\documentclass[twocolumn]{aastex631}

\newcommand\nickel{$^{56}$Ni }

\accepted{\today}

\submitjournal{ApJ}

\usepackage{amsmath}

\shorttitle{}
\shortauthors{Sawada, Kashiyama and Suwa}

\begin{document}


\title{On the energy source of ultra-stripped supernovae}

\correspondingauthor{Ryo Sawada}
\email{ryo@g.ecc.u-tokyo.ac.jp}

\author[0000-0003-4876-5996]{Ryo Sawada}
\affiliation{Department of Earth Science and Astronomy, Graduate School of Arts and Sciences, The University of Tokyo, Tokyo 153-8902, Japan}
\affiliation{Department of Astrophysics and Atmospheric Sciences, Faculty of Science, Kyoto Sangyo University, Kyoto 603-8555, Japan}

\author[0000-0003-4299-8799]{Kazumi Kashiyama}
\affiliation{Department of Physics, Graduate School of Science, The University of Tokyo, Tokyo 113-0033, Japan}
\affiliation{Research Center for the Early Universe, Graduate School of Science, The University of Tokyo, Tokyo 113-0033, Japan}
\affiliation{Kavli Institute for the Physics and Mathematics of the Universe (Kavli IPMU,WPI), The University of Tokyo, Chiba 277-8582, Japan}

\author[0000-0002-7443-2215]{Yudai Suwa}
\affiliation{Department of Earth Science and Astronomy, Graduate School of Arts and Sciences, The University of Tokyo, Tokyo 153-8902, Japan}
\affiliation{Department of Astrophysics and Atmospheric Sciences, Faculty of Science, Kyoto Sangyo University, Kyoto 603-8555, Japan}
\affiliation{Center for Gravitational Physics, Yukawa Institute for Theoretical Physics, Kyoto University, Kyoto 606-8502, Japan}

\begin{abstract}
Ultra-stripped supernovae (USSN) with a relatively low ejecta mass of $\sim0.1M_\odot$ (e.g., iPTF14gqr and SN2019dge) are considered to originate from ultra-stripped carbon-oxygen (CO) cores in close binary systems and are likely to be progenitors of binary neutron stars (BNSs). Here we conduct the explosion simulations of ultrastripped progenitors with various masses ($1.45\,M_\odot \leq M_\mathrm{CO} \leq 2.0\,M_\odot$) based on results of neutrino-radiation hydrodynamics simulations, and consistently calculate the nucleosynthesis and the SN light curves. We find that a USSN from a more massive progenitor has a larger ejecta mass but a smaller $^{56}$Ni mass mainly due to the fallback, which leads to the light curve being dimmer and slower. By comparing the synthetic light curves with the observed ones, we show that SN2019dge can be solely powered by $^{56}$Ni synthesized during the explosion of a progenitor with $M_\mathrm{CO} \lesssim 1.6\,M_\odot$ while iPTF14gqr cannot be explained by the $^{56}$Ni powered model; $\sim 0.05M_\odot$ of $^{56}$Ni inferred from the light curve fitting is argued to be difficult to synthesize for ultra-stripped progenitors. We consider fallback accretion onto and rotation-powered relativistic wind from the newborn NS as alternative energy sources and show that iPTF14gqr could be powered by a newborn NS with a magnetic field of $B_p \sim 10^{15}\,\mathrm{G}$ and an initial rotation period of $P_i \sim 0.1\,\mathrm{s}$.

\end{abstract}

\section{Introduction}\label{sec:intro}

The progenitor system of binary neutron stars (BNSs) is still under debate. The most likely candidate is ultra-stripped supernova (USSN), which has typically ten times smaller ejecta mass ($\mathcal{O}(0.1)M_\odot$) than a canonical supernova \citep{2015MNRAS.451.2123T}. 
Because the close binary interactions tear off the envelope of the secondary star in a binary system, the second supernova in the system would have little ejecta mass. The small ejecta mass also helps to prevent the binary system from breaking up.
Recent progress of transient surveys has led to successful observations of USSNe from the early phase just after the explosion. In particular, \cite{2018Sci...362..201D} observed iPTF 14gqr in the shock-cooling phase, indicating the presence of the extended envelope with $500 R_\odot$ and $0.01M_\odot$. This result implies that USSN is of massive star origin.

The energy source of USSN is assumed to be radioactive \nickel as in the case of canonical stripped envelope SNe. From the light curve fitting, the amount of \nickel in USSN ejecta is estimated to be 0.05$M_\odot$ for SN 2005ek \citep{2013ApJ...778L..23T}, 0.05$M_\odot$ for iPTF 14gqr \citep{2018Sci...362..201D}, and 0.015$M_\odot$ for SN2019dge \citep{2020ApJ...900...46Y}. These values are slightly smaller than those inferred for canonical SNe \cite[see, e.g.,][]{2020A&A...641A.177M}, which may be consistent with the explosion energy of USSNe estimated to be several times to an order of magnitude smaller than that of canonical SNe. 

In order to theoretically calculate the \nickel amount loaded on the SN ejecta, we need detailed hydrodynamics simulations and nucleosynthesis calculations. 
A lot of these studies have been done for canonical SNe
\cite[and references therein]{2018SSRv..214...62T}.
Although not as systematically investigated as canonical SNe, some previous works have performed calculations on nucleosynthesis in USSNe. 
Based on the first neutrino-radiation hydrodynamics simulation of USSNe by \cite{2015MNRAS.454.3073S}, \cite{2017MNRAS.471.4275Y} conducted the detailed nucleosynthesis calculations for two cases. \cite{2018MNRAS.479.3675M} followed a similar simulation up to the shock breakout of a USSN progenitor star and investigated the explosive nucleosynthesis.
The explosion energy of these simulations ($\sim 10^{50}$ erg) was consistent with the values obtained from the light curve fitting, but the \nickel amount ($\sim$ 0.01$M_\odot$) was insufficient. 
Additionally, \cite{2017MNRAS.466.2085M} have calculated the explosive nucleosynthesis, light curve and spectrum of USSNe with a simplified explosion simulation.
Still the synthesized amount of \nickel in the simulated USSNe is approximately $\sim 0.03 M_\odot$, and not sufficient for SN 2005ek and iPTF 14gqr.
We recall this problem {\it \nickel problem} here.
It should be noted that all the above studies on nucleosynthesis in USSNe were investigated in a limited number of progenitor models. A systematic study is needed to make a detailed comparison between the theory and observation in USSNe.

The \nickel problem has recently been discussed as an inherent problem not only in USSNe but also in canonical SN explosion simulations \citep{2019MNRAS.483.3607S,2019ApJ...886...47S,2021ApJ...908....6S}. It has long been known that the explosion energies obtained in simulations are significantly lower than the observed typical values. Recently, updated simulations have been reported to reach $10^{51}$ erg \citep{2021ApJ...915...28B,2021Natur.589...29B}, and the explosion energy problem is gradually being solved. It should be noted, however, that almost all other simulations have not yet been able to reproduce it (see e.g., \citealt{2012ARNPS..62..407J,2016MNRAS.461L.112T,2021Natur.589...29B} and references therein). Additionally, these simulations have yet to produce a sufficient amount of $^{56}$Ni. The reason for this is that the explosive nucleosynthesis requires a rapid increase in the explosion energy to produce \nickel, whereas the simulation results show a slow increase.

In USSNe, such explosion energy problem does not exist. 
Therefore, \nickel production in USSNe can be discussed more robustly than in normal supernovae.
In this study, we perform a systematic study of the energy sources of USSN light curves with a wider range of progenitor models, connecting long-term hydrodynamics simulations with nucleosynthesis calculations to light curve calculations.
We first simulate one-dimensional hydrodynamics and nucleosynthesis, with the explosion model  reconstructed from the results of neutrino-radiation hydrodynamic simulations \citep{2015MNRAS.454.3073S}.
We then compare the numerical results by calculating the light curves of the observed transients iPTF 14gqr and SN 2019dge using the analytical model.

This paper is organized as follows. 
We describe our model to solve the core-collapse explosion of ultra-stripped progenitors and calculate the nucleosynthesis and the SN light curves in Sec. \ref{sec:model}. 
We show the results of our calculation and compare the \nickel-powered light curves with the observed USSNe in Sec. \ref{sec:result}. 
We discuss the uncertainties in our model and consider alternative energy sources of USSNe in Sec. \ref{sec:discus}. We conclude the paper in Sec. \ref{sec:concl}.

\section{Model and Method}\label{sec:model}

\subsection{Progenitors}\label{sec:progenitor}

We use the 7 pre-explosion models of CO stars same as \citet{2015MNRAS.454.3073S}. These models are computed from an initial CO core of mass 1.45-2 $M_\odot$ (1.45, 1.5, 1.6, 1.7, 1.8, 1.9 and 2.0 $M_\odot$) at the central C burning using the stellar evolution code described in \cite{2015MNRAS.454.3073S,2017MNRAS.471.4275Y}. We will refer to CO core models with a mass of x.yz $M_\odot$ as COxyz model. These CO cores correspond to the secondary stars in the close binary system and are expected to lose their H and He-rich envelopes during the binary evolution with the primary NS. See also \cite{2017MNRAS.471.4275Y,2018MNRAS.481.3305S} for details. 

\begin{figure}
\centering
  \includegraphics[width=0.475\textwidth]{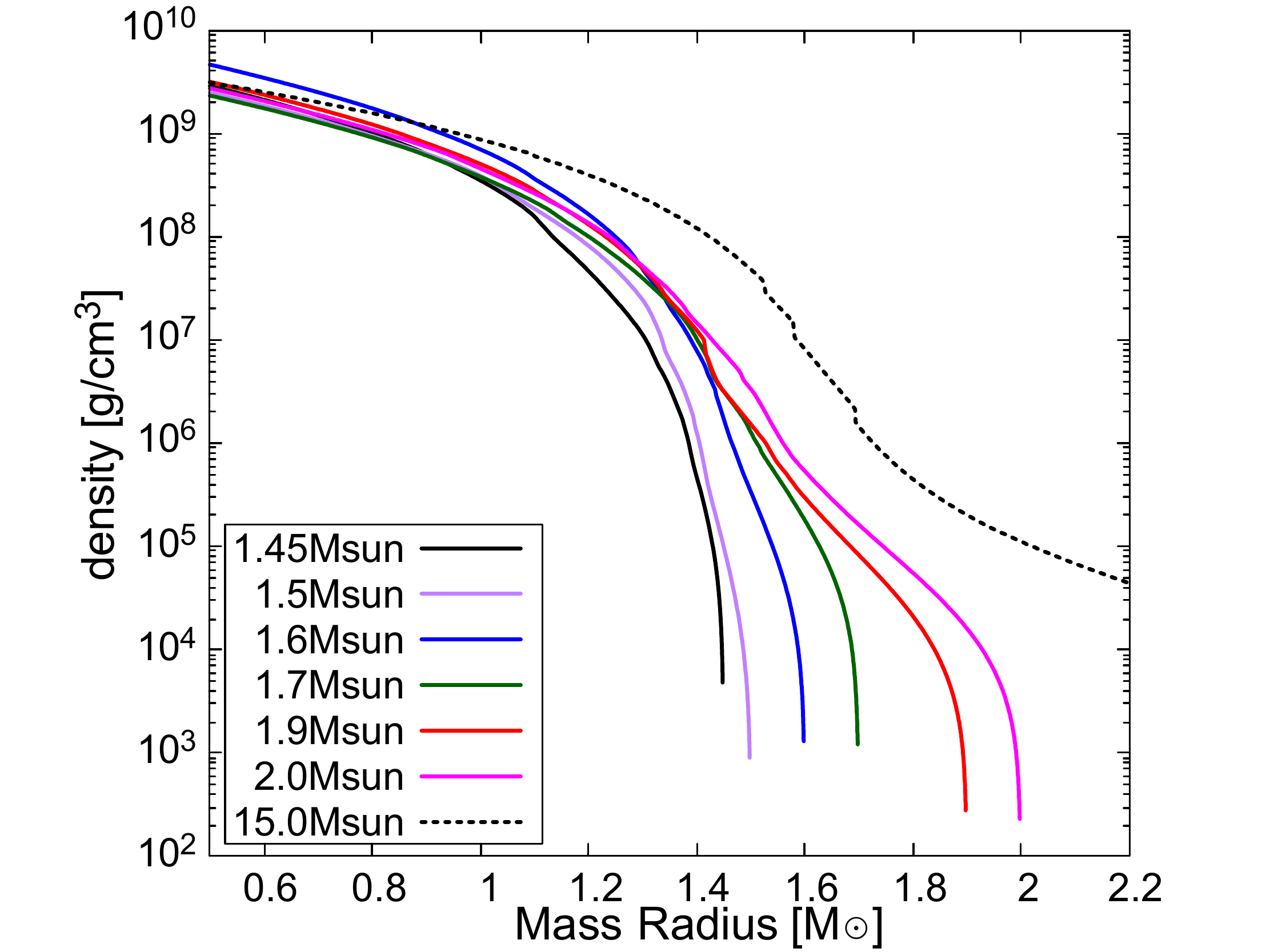} 
  \includegraphics[width=0.475\textwidth]{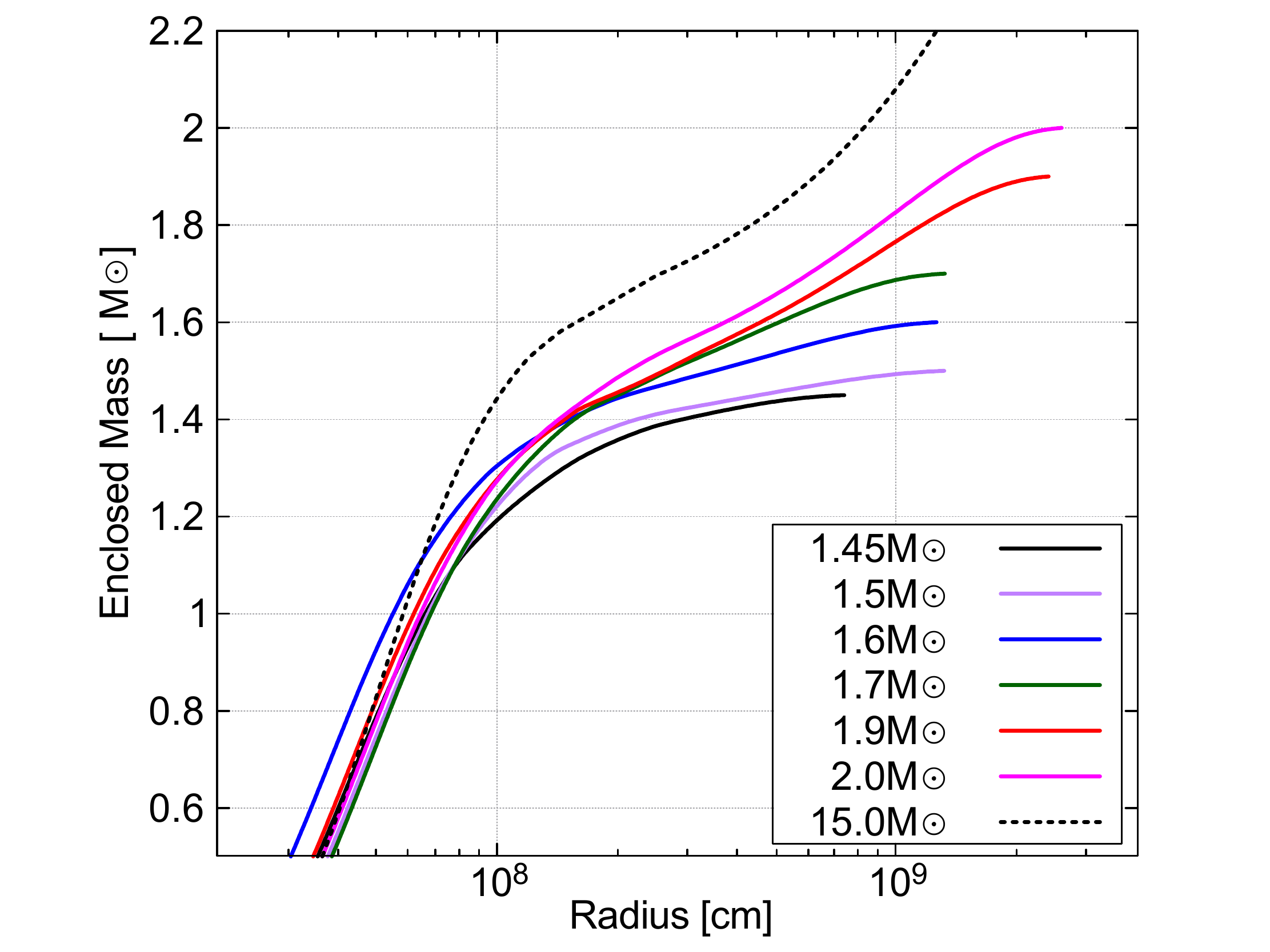} 
\caption{Pre-collapse structure of ultra-stripped supernova progenitors with masses of $1.45-2.0 \,M_\odot$. The top and bottom panels show enclosed mass vs density and radius vs enclosed mass, respectively. The dotted line shows the pre-collapse structure of a canonical supernova progenitor with a zero-age-main-sequence mass of $15.0\,M_\odot$.} 
\label{fig:preSN}
\end{figure}

Figure \ref{fig:preSN} shows the enclosed-mass profiles of the USSN progenitor model. For comparison, we also show a canonical-SN progenitor model by the dotted line with ZAMS mass of $15.0 M_\odot$ obtained by the stellar evolution code {\tt MESA} \citep{2015ApJS..220...15P}. 
This figure shows that USSN progenitors have smaller core masses than the 15$M_\odot$ model. For example, at $\rho=10^6$ g cm$^{-3}$, USSN progenitor masses are 1.4--1.6 $M_\odot$, whereas the 15$M_\odot$ model is 1.8 $M_\odot$. This result is attributed to the difference in core entropy, as shown in \citet{2018MNRAS.481.3305S}: the USSN progenitor has smaller core entropy than the 15$M_\odot$ model due to the absence of entropy inflow from the high-entropy envelope. As a result, the increase in Chandrasekhar mass due to the finite temperature effect is ineffective, resulting in a smaller core mass when gravitational collapse occurs.
Such a relatively small core mass has been reproduced in recent stellar evolution calculations including binary interaction effects \citep{2021ApJ...920L..36J}.

\begin{table*}
\centering
    \caption{Properties of the explosion models and summary of simulation results.}\label{tbl:USSN}
    \begin{tabular}{ r  c  c c | c c ||  c c c c c } \hline \hline
    &\multicolumn{3}{c|}{ Our work} &
    \multicolumn{2}{c||}{\citet{2015MNRAS.454.3073S}} &
    \multicolumn{5}{c}{ Our result}  \\\hline 
      Model & $M_\mathrm{CO}$ 
      & $E_\mathrm{expl}$ & $M_\mathrm{PNS,i}$ 
      & $E_\mathrm{expl}$ & $M_\mathrm{PNS}$ 
      & $E_\mathrm{expl}$ & $M_\mathrm{ej}$ 
      & $M_\mathrm{Ni}$ 
      & ${M}_\mathrm{fb}$ & $M_\mathrm{NS,f}$ \\
      &  ($M_\odot$)& (Bethe) & ($M_\odot$) & (Bethe) & ($M_\odot$) & (Bethe) & ($M_\odot$) & ($M_\odot$) 
      & ($10^{-2}M_\odot$) & ($M_\odot$)\\ \hline 
      CO145 & 1.45 & 0.17 & 1.35 & 0.177 & 1.35 
      & 0.17 & 0.097 & $1.63\times10^{-2}$ & $0.30$  & 1.35 \\
      CO15 & 1.5& 0.15 & 1.36 & 0.153 & 1.36
      & 0.15 & 0.136 & $1.38\times10^{-2}$ & $0.54$  & 1.37 \\
      CO16 & 1.6& 0.12 & 1.42 & 0.124 & 1.42 
      & 0.12 & 0.151 & $1.20\times10^{-2}$ & $0.56$  & 1.43 \\
      CO17 & 1.7& 0.12 & 1.45 & - & -  
      & 0.12 & 0.225 & $1.09\times10^{-2}$ & $1.04$  & 1.46 \\
      CO18 & 1.8& 0.12 & 1.49 & 0.120 & 1.49  
      & 0.12 & 0.277 & $4.80\times10^{-4}$ & $1.23$  & 1.50 \\
      CO19 & 1.9& 0.12 & 1.54 & - & -  
      & 0.12 & 0.307 & $9.19\times10^{-5}$ & $2.11$  & 1.56 \\
      CO20 & 2.0& 0.12 & 1.60 & - & 1.60 
      & 0.12 & 0.286 & $7.78\times10^{-5}$ & $4.02$  & 1.64 \\
      \hline \hline
    \end{tabular}
\end{table*}


\subsection{Core-collapse explosion}\label{sec:hydro}

In order to estimate the \nickel amount loaded on the SN ejecta, one has to solve the long-term evolution including the fallback effect. 
Since the neutrino-radiation hydrodynamics simulations like \citet{2015MNRAS.454.3073S} are not suitable for 1000-second long simulations because of its computational cost and numerical limitation of tabulated equation of state, we adopt a different approach in this work.
For a given progenitor, we trigger an explosion by instantaneously injecting an amount of thermal energy ($E_\mathrm{inject}$) at a certain mass radius ($M_\mathrm{inject}$). 
In our calculations, the initial proto-NS mass $M_\mathrm{PNS, i}$ is defined as the total mass of matter that has never been ejected outside a radius of $R_c=10^{10}$ cm, where the escape velocity becomes small enough compared to the ejecta velocity.
We repeat the simulation with varying ($E_\mathrm{inject}$, $M_\mathrm{inject}$) until both the explosion energy $E_\mathrm{expl}$ and the initial proto-NS mass $M_\mathrm{PNS, i}$ become consistent with those of \citet{2015MNRAS.454.3073S}. See Table \ref{tbl:USSN}.

We note that different pairs of ($E_\mathrm{inject}$, $M_\mathrm{inject}$) can give a same ($E_\mathrm{expl}$,$M_\mathrm{PNS, i}$) even for a given progenitor model. In this case, we employ the case with a sufficiently small $M_\mathrm{inject}$ as our physical model, for which the properties of the explosion converges. We also note that no calculations were performed in \citet{2015MNRAS.454.3073S} for some progenitor models, where we interpolated $M_\mathrm{PNS,i}$ from the results in CO star model with close mass. In addition, we adopted $E_\mathrm{expl}= 0.12$ Bethe (1 Bethe=$1\times10^{51}$ erg) for all CO star models heavier than the CO16 model for simplicity, since the explosion energy tended to converge at $\sim0.12$ Bethe for those CO star models in \citet{2015MNRAS.454.3073S}. We will discuss impacts of these assumptions in section \ref{sec:nickel}. 

For the spherically symmetric simulation, we use an 1D Euler code based on {\tt hydro1d},\footnote{http://zingale.github.io/hydro1d/} which employs a Godunov-type scheme to integrate the conservation equations with a gravity source term; 
\begin{align}
    &\cfrac{\partial \rho}{\partial t} + \cfrac{1}{r^2}\cfrac{\partial}{\partial r} \left(r^2\rho v_r\right) =0 \label{eq:basic1}~,\\
    &\cfrac{\partial v_r}{\partial t} + v_r\cfrac{\partial v_r}{\partial r} =  -\cfrac{1}{\rho}\cfrac{\partial P}{\partial r} - \cfrac{GM(r,t)}{r^2}
    \label{eq:basic2}~,\\
    &\cfrac{\partial( \rho e ) }{\partial t}+ \cfrac{1}{r^2}\cfrac{\partial [(\rho e + P)v_r r^2] }{\partial r} =- \rho v_r\cfrac{GM(r,t)}{r^2} \label{eq:basic3}~,
\end{align}
where $\rho$, $v_r$, $P$, $e$, and $M(r,t)$ are density, radial velocity, total pressure, specific energy, and enclosed mass, respectively. The enclosed mass is the sum of the mass inside the inner boundary $M_G(t)$ and the mass in the computational domain interior to the radius $r$, i.e., 
\begin{equation}\label{eq:basic5}
M(r,t) = M_G(t) + 4\pi\int_{R_{\rm in}}^r \rho(r',t)r'^2 dr'.
\end{equation}
The mass flux flowing through the inner boundary of the computational domain is added to $M_G(t)$. We employ the Helmholtz equation of state \citep{2000ApJS..126..501T} and neglect the effects of weak interaction and nuclear burning on the dynamics. 

We employ a logarithmic grid with 300 mesh per decade in radius, i.e., the grid size ratio is fixed to be $\Delta r/r \approx 0.008$. For $t \leq 10\,\mathrm{sec}$, we fix the inner boundary at a mass radius 0.1 $M_\odot$ inner from the initial proto-NS mass radius and the outer boundary at twice the progenitor radius. Here $t=0\,\mathrm{sec}$ corresponds to the onset of the explosion. For $t \geq 10\,\mathrm{sec}$, we move both the inner and outer boundary radii outwards as the SN ejecta expands; every time the forward shock arrives at 85 \% of the outer boundary radii, we reset the inner boundary $R_\mathrm{in} \rightarrow R_\mathrm{in} + \Delta R_\mathrm{in}$ so that $\Delta M_G \leq 10^{-3}\, M_\odot$ and $\Delta R_\mathrm{in} \leq 5\,R_\mathrm{in}$ and the new outer boundary is set to be at $R_\mathrm{out} = 10^5\,R_\mathrm{in}$ with keeping $\Delta r/r \approx 0.008$. All the primitive variables are remapped to the new mesh. We continue the calculation up to $t \sim 10\,\mathrm{day}$. Note that we maintain the overall mass conservation within the machine precision.

\subsection{Nucleosynthesis}
The nucleosynthesis calculations are performed as a post-processing analysis.
We distribute tracer particles with a fixed mass of $m=10^{-4}M_\odot$ to fill the computational domain of the the  spherically symmetric simulation. 
Each particle captures the time evolution of the hydrodynamic quantities, with which we calculate a reaction network of 640 nuclear species with the \texttt{torch}\footnote{http://cococubed.asu.edu/} code \citep{1999ApJS..124..241T}.

\subsection{Supernova light curve}\label{sec:lc}
We calculate the bolometric light curves of the USSNe by the one-zone model~\citep{1980ApJ...237..541A,1982ApJ...253..785A,2010ApJ...717..245K,2013ApJ...772...30D} with using the $^{56}$Ni mass $M_\mathrm{Ni}$, the ejecta mass $M_\mathrm{ej}$, and the explosion energy $E_\mathrm{expl}$ obtained by the core-collapse explosion simulations as the input parameters. The energy conservation of the SN ejecta is described as 
\begin{align}
    \cfrac{\partial E_\mathrm{int}(t)}{\partial t} &= -P\cfrac{\partial V(t)}{\partial t} +\dot{Q}(t) - L(t)~, \label{eq:ap1}\\
    \cfrac{\partial E_\mathrm{kin}(t)}{\partial t} &= P\cfrac{\partial V(t)}{\partial t} ~, \label{eq:ap2}
\end{align}
where $E_\mathrm{int}(t)$ is the internal energy,  $E_\mathrm{kin}(t)$ is the kinetic energy, $\dot{Q}(t)$ is the heating rate by some energy injection processes, and $L(t)$ is the radiative cooling rate. The first term in the right hand side of Equation (\ref{eq:ap1}) represents the adiabatic loss of the internal energy, which is converted into the kinetic energy through Equation (\ref{eq:ap2}). The expansion velocity of the ejecta is given as  
\begin{equation}
    v_\mathrm{ej}(t) = \sqrt{2 E_\mathrm{kin}(t)/ M_\mathrm{ej}}~, \label{eq:ap3}
\end{equation}
and the radius evolves as
\begin{equation}
    R_\mathrm{ej}(t) = \int_0^t v_\mathrm{ej}(t') dt'~. \label{eq:ap4}
\end{equation}
Assuming that the internal energy is dominated by radiation, the adiabatic loss term can be rewritten as
\begin{equation}\label{eq:ad_loss}
    P\cfrac{\partial V(t)}{\partial t} 
    =\cfrac{E_\mathrm{int}(t)}{R_\mathrm{ej}(t)} v_\mathrm{ej}(t)
    =\cfrac{E_\mathrm{int}(t)}{t_\mathrm{dyn}},
\end{equation}
where
\begin{equation}
    t_\mathrm{dyn}=\cfrac{R_\mathrm{ej}(t) }{v_\mathrm{ej}(t) }~\label{eq:ap6}
\end{equation}
is the dynamical timescale of the ejecta. On the other hand, the radiative loss term can be written as 
\begin{equation}\label{eq:rad_loss}
    L(t)
    =\cfrac{E_\mathrm{int}  t_\mathrm{dyn}}{t_\mathrm{diff}^2}, 
\end{equation}
where 
\begin{equation}\label{eq:ap10} 
    t_\mathrm{diff}
    =\left[\cfrac{3}{4\pi}\cfrac{\kappa M_\mathrm{ej}}{v_\mathrm{ej}c}\cfrac{1}{\xi}\right]^{1/2}
\end{equation}
is the diffusion timescale $t_\mathrm{diff}$ through the ejecta, $\kappa = 0.07\,\mathrm{cm^2\,g^{-1}}$ is the opacity \citep[e.g.,][]{2000ApJ...530..757P,2018A&A...609A.136T}, and $\xi=\pi^2/3$ represents the geometrical factor~\citep{1982ApJ...253..785A}. Substituting Eqs. (\ref{eq:ad_loss}) and (\ref{eq:rad_loss}) into Eqs. (\ref{eq:ap1}) and (\ref{eq:ap2}),
\begin{align}
    \cfrac{\partial E_\mathrm{int}(t)}{\partial t} 
    &= -\cfrac{ E_\mathrm{int}(t)}{t_\mathrm{dyn}}  +\dot{Q}(t) 
    - \cfrac{E_\mathrm{int}  t_\mathrm{dyn}}{t_\mathrm{diff}^2} ~,\label{eq:ap12}\\
    \cfrac{\partial E_\mathrm{kin}(t)}{\partial t} 
    &= \cfrac{ E_\mathrm{int}(t)}{t_\mathrm{dyn}} ~.\label{eq:ap13}
\end{align}
For a given heating rate $\dot Q(t)$, we solve Eqs. (\ref{eq:ap12}) and (\ref{eq:ap13}) with the 4th-order Runge-Kutta method to obtain $E_\mathrm{int}(t)$ and $E_\mathrm{kin}(t)$, and then the SN light curve $L(t)$.

By default, we consider the radioactive decay of $^{56}$Ni as the main heating source of the ejecta. In this case, 
\begin{align}
    \dot{Q}(t) &= f_\mathrm{dep}\cdot
    \left(  M_\mathrm{Ni} \,\ q_\mathrm{Ni}(t) \right)\\
    q_\mathrm{Ni}(t) &= \epsilon_\mathrm{Ni}\cdot e^{-t/\tau_\mathrm{Ni}}
    +\epsilon_\mathrm{Co}\cdot e^{-t/\tau_\mathrm{Co}}  \label{eq:ap14}~,
\end{align}
where $\epsilon_\mathrm{Ni}=3.22\times 10^{10}$ erg g$^{-1}$ s$^{-1}$ and $\epsilon_\mathrm{Co}=6.78\times 10^{9}$ erg g$^{-1}$ s$^{-1}$ are the specific decay energy of $^{56}$Ni and $^{56}$Co, and $\tau_\mathrm{Ni}=8.8$ day and $\tau_\mathrm{Co}=113.6$ day are the mean lifetimes of $^{56}$Ni and $^{56}$Co, respectively. We consider alternative energy sources in Sec. \ref{sec:obs}.

\section{Results}\label{sec:result}

\begin{figure}
    \centering
    \includegraphics[width=0.475\textwidth]{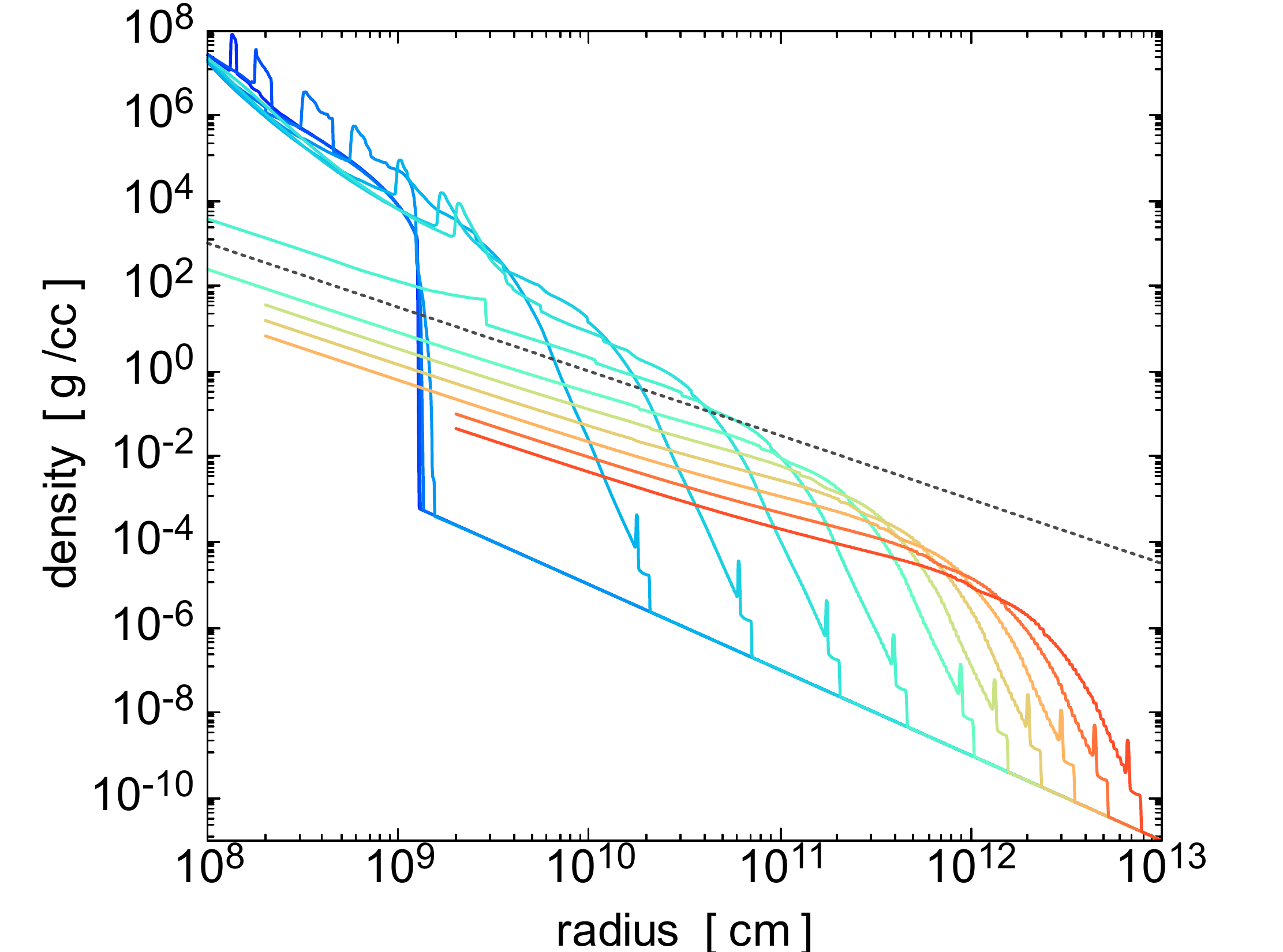} 
    \caption{Density profiles of the core-collapse explosion of the CO145 model from $t = 10\,\mathrm{msec}$ to $t \approx 10^4\,\mathrm{sec}$. The dashed line indicates the $r^{-3/2}$ relation.
    }
    \label{fig:hydro}
\end{figure}

Figure \ref{fig:hydro} shows snapshots of the density profile in the CO145 model.
For numerical reasons, we fill the outer region of the star with an ambient medium in hydrostatic equilibrium with $\rho \propto r^{-2}$.
The ambient density is chosen as $\rho (r)=10^{15} (r/1\,{\rm cm})^{-2}$ g cm$^{-3}$, and the total mass in the computational domain is less than $10^{-4} M_\odot$ to avoid artificially slowing down the ejecta.
While the structure of outside the star in the low-density ambient medium might influence the forward shock velocity, it does not significantly affect the dynamics of the inner fallback region \citep[see e.g.,][]{2018MNRAS.476.2366F}.
The inner region has a structure as $\rho \propto r^{-3/2}$, which is derived by the approximately constant mass accretion rate, $\dot M\propto \rho r^2v_{\rm ff}$ with the free-fall velocity $v_{\rm ff}\propto r^{-1/2}$, 
when the shock reaches $10^{12}$ cm, corresponding to about $t\approx1000$ sec after the explosion.
Here, we should note that, in the previous core-collapse explosoin calculations for progenitors with massive outer layer, the inward reverse shock self-reflecting at the inner boundary was observed~\citep{2016ApJ...821...69E}. 
The timing of the self-reflection is sensitive to the position of the inner boundary and artificially alter the fallback dynamics~\citep{2021ApJ...920L..17V}.
In cases of USSN progenitors without massive outer layer, there is no reverse shock causing the above self-reflection problem.

\begin{figure}
\centering
  \includegraphics[width=0.475\textwidth]{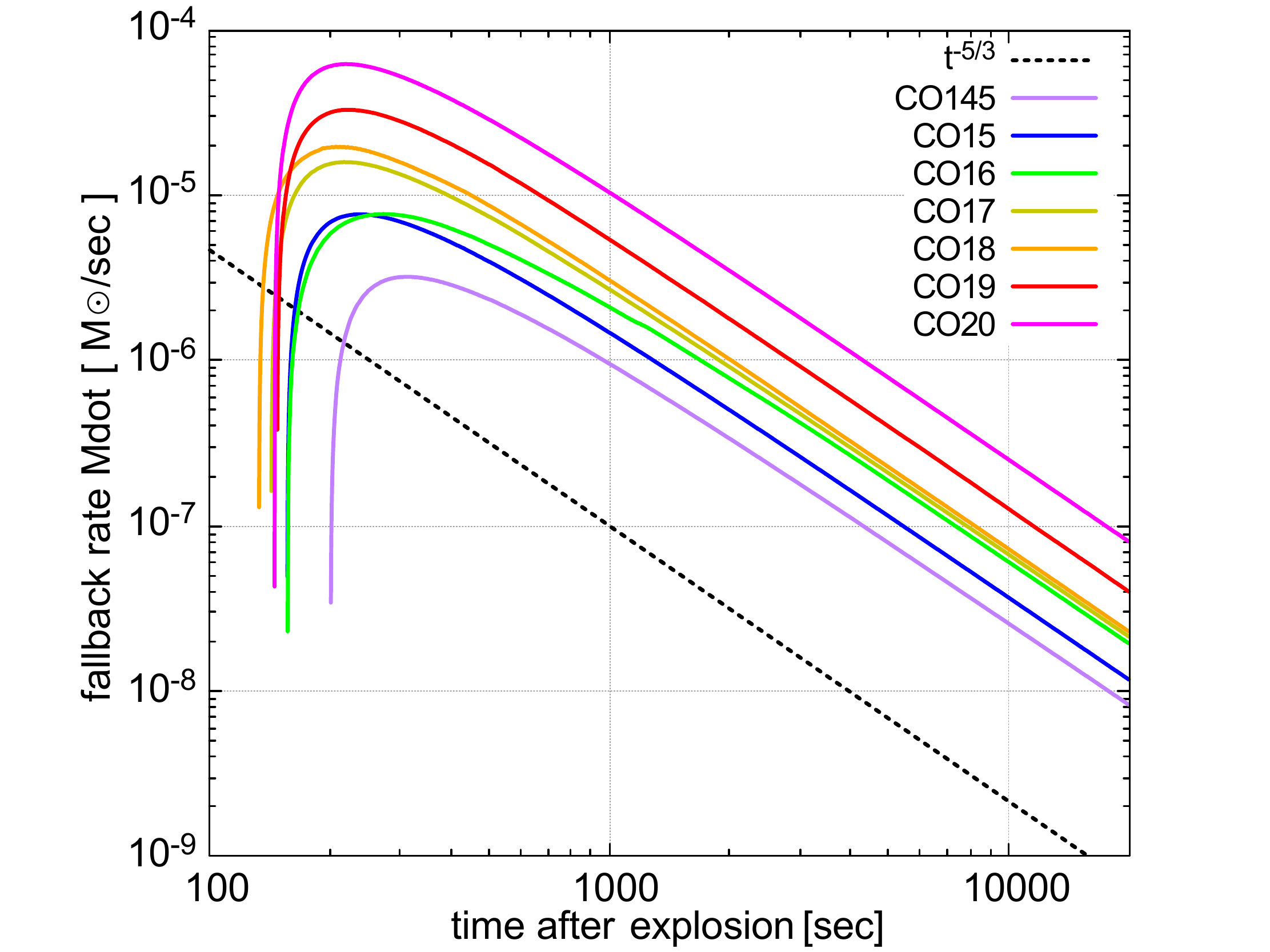} 
\caption{Time evolution of the mass accretion rate evaluated at $r=10^{10}$ cm, where t = 0 corresponds to the onset of the explosion.} 
\label{fig:fbrate}
\end{figure}

Figure \ref{fig:fbrate} shows the time evolution of the SN fallback. We estimate the mass accretion rate as  
\begin{equation}
\dot{M}_\mathrm{fb}(t) = 4\pi R_\mathrm{c}^2 \rho_\mathrm{c}(t) v_{r,\mathrm{c}}(t) \label{eq:basic6}~,
\end{equation}
where $\rho_\mathrm{c}$ and $v_{r,\mathrm{c}}$ are density and radial velocity at radius $R_\mathrm{c}=10^{10}$ cm, respectively. We find a larger fallback mass for a larger CO core mass case; the lightest CO145 model has a total fallback mass of $M_\mathrm{fb}\lesssim 0.01\,M_\odot$ while the heaviest CO20 model has $M_\mathrm{fb}\sim 0.04\,M_\odot$. 
These values are comparable to those estimated for canonical-SN explosions \citep[e.g.,][]{2008ApJ...679..639Z,2021arXiv210407493J}; the smaller explosion energy of USSNe than canonical SNe is compensated by the less dense core of ultra-stripped progenitors to give a comparable fallback mass. In all models, the fallback rate declines as $\propto t^{-5/3}$ at $t\gtrsim1000$ sec \citep{1988Natur.333..644M,1989ApJ...346..847C}. We note that, in the case of USSNe, there is no enhancement of the fallback rate in a later phase that could occur for a progenitor with a helium layer.

\begin{figure}
\centering
  \includegraphics[width=0.475\textwidth]{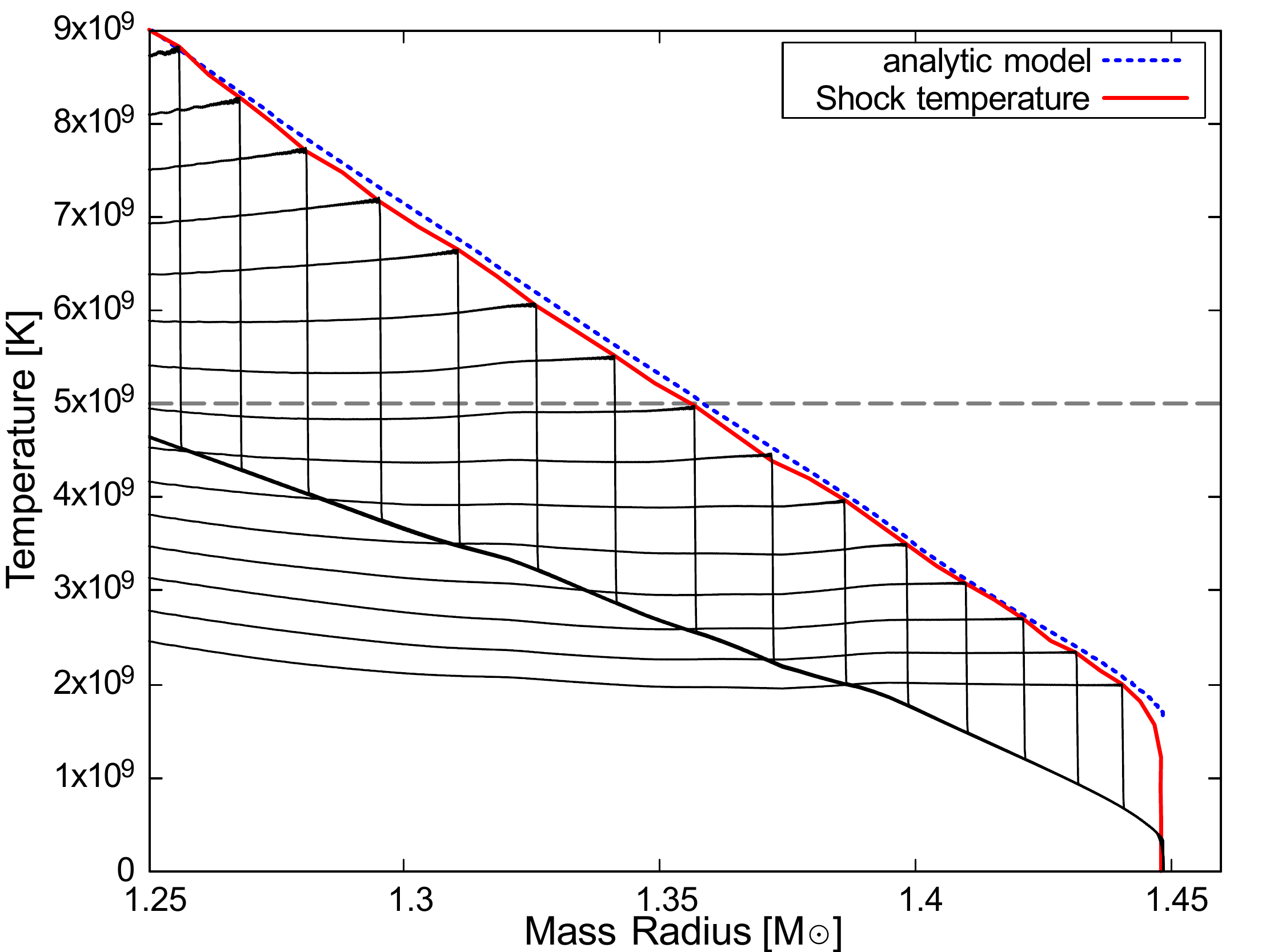} 
\caption{Temperature profiles of the core-collapse explosion of the CO145 model with respect to mass radius $M_r$. The black lines show the snapshots of every $10\,\mathrm{ms}$ from $t = 10\,\mathrm{msec}$ to $t = 150\,\mathrm{msec}$. The red line marks the time evolution of the temperature at the shock front while the dashed blue line indicates the analytical model given by Equation \eqref{eq:app1}.}
\label{fig:CO145}
\end{figure}

Figure \ref{fig:CO145} shows the time evolution of the temperature profile of the CO145 model.
The temperature at the shock front can be well presented by an analytical expression \citep{2002RvMP...74.1015W} as
\begin{equation}
    T_\mathrm{shock}(r) = 7.48\times 10^9 \mathrm{~K~}
    \left(\cfrac{E(r)}{10^{50}\mathrm{erg}}\right)^{1/4}
    \left(\cfrac{r}{10^{8}\mathrm{cm}}\right)^{-3/4}~,\label{eq:app1}
\end{equation}
where $E(r) = E_\mathrm{expl} + |E_\mathrm{bind}(r)|$ is an energy in the post shock region, and $|E_\mathrm{bind}(r)|$ is the binding energy of the progenitor outside the radius $r$.
In regions where the postshock temperature exceeds $5\times10^9$ K, the nuclear statistical equilibrium is realized within a dynamical timescale,
and the material turns into iron group elements, mainly \nickel~\citep[e.g.,][]{1996ApJ...460..408T,2013ARA&A..51..457N}.
The critical radius for the nucleosynthesis can be estimated from Equation \eqref{eq:app1} as
\begin{equation}
    R_{T_9=5} = 2.17\times 10^8 \mathrm{~cm~}
    \left(\cfrac{E(r)}{10^{50}\mathrm{erg}}\right)^{1/3}~,\label{eq:app2}
\end{equation}
where $T_9 = T/10^9$ K.

\begin{figure}
\centering
  \includegraphics[width=0.475\textwidth]{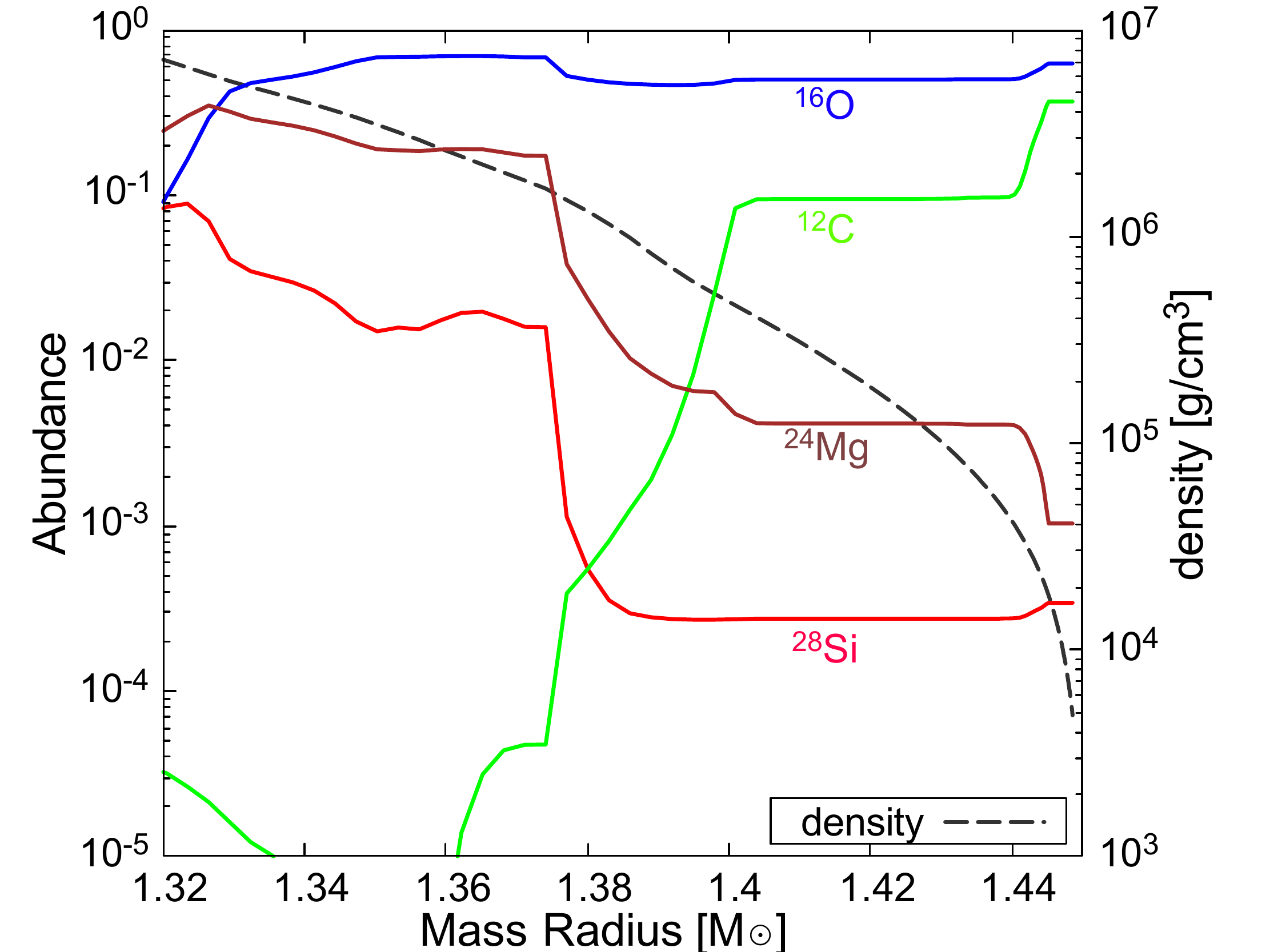} 
  \includegraphics[width=0.475\textwidth]{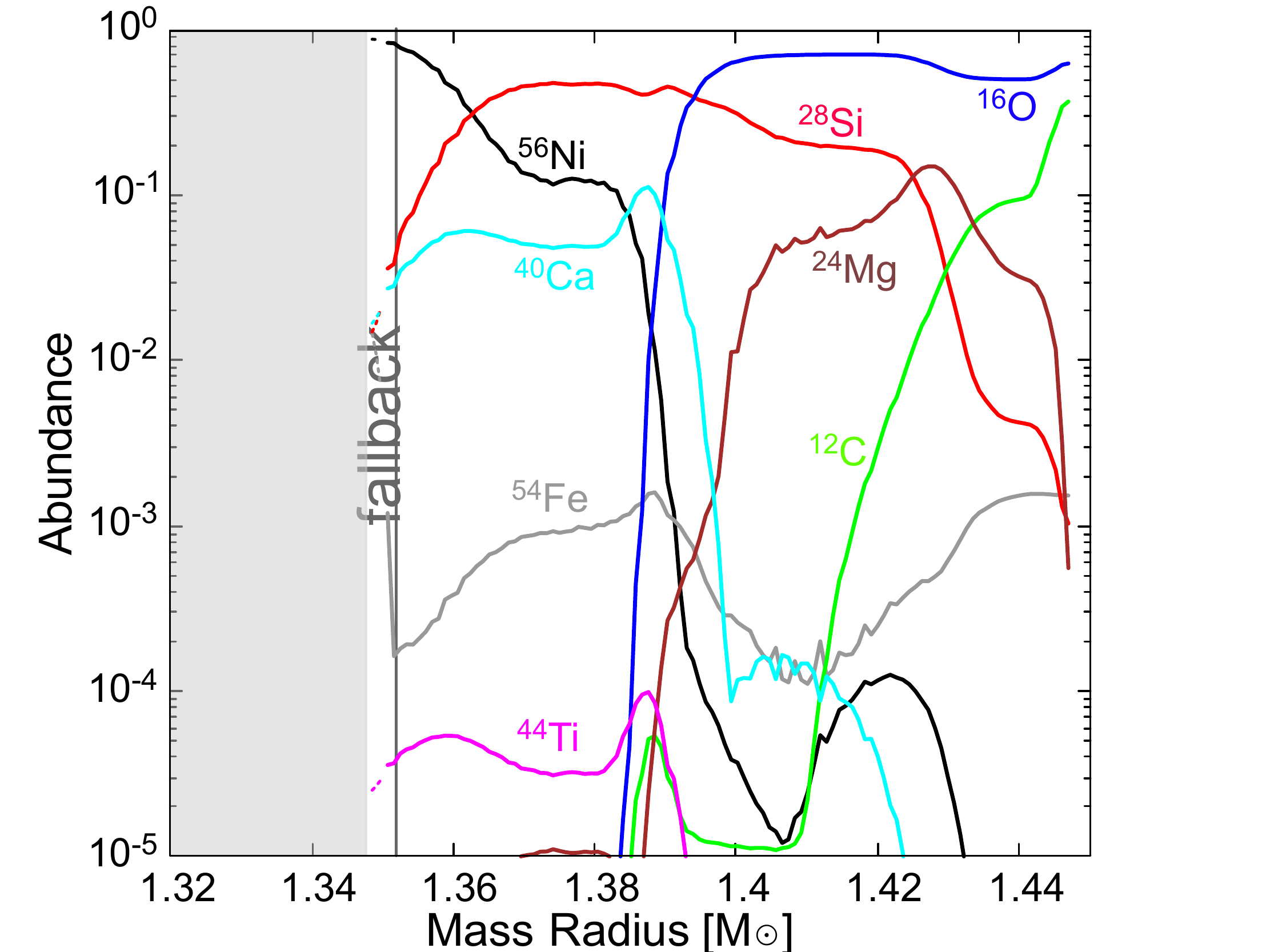} 
\caption{
Explosive nucleosynthesis of the CO 145 model. The top panel shows the metal abundance with respect to the mass radius in the pre-collapse phase. The bottom panel shows the results of the core-collapse explosion. The vertical line separates the supernova ejecta and the fallback matter. 
} 
\label{fig:CO145nuc}
\end{figure}

\begin{figure}
\centering
  \includegraphics[width=0.475\textwidth]{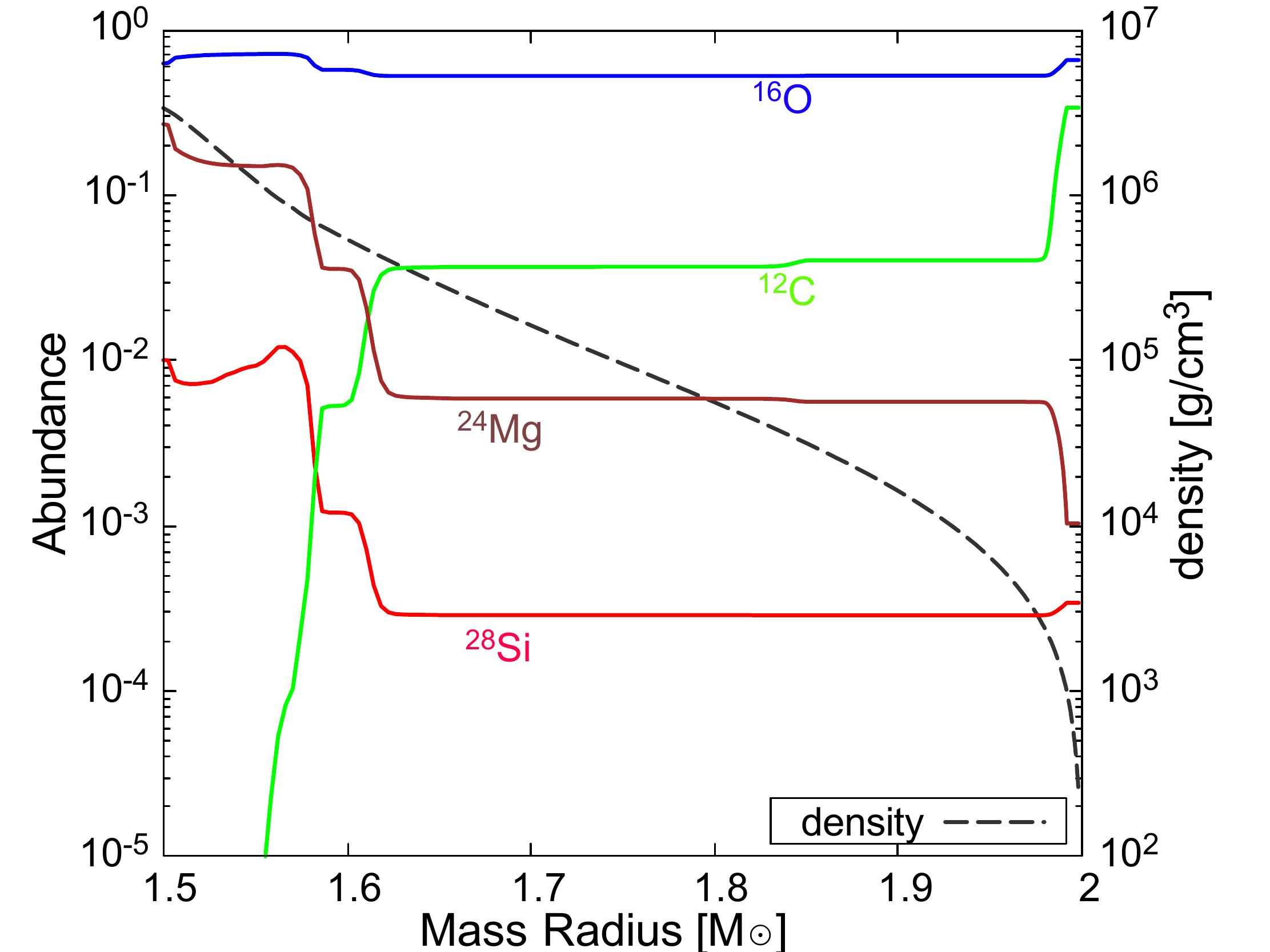} 
  \includegraphics[width=0.475\textwidth]{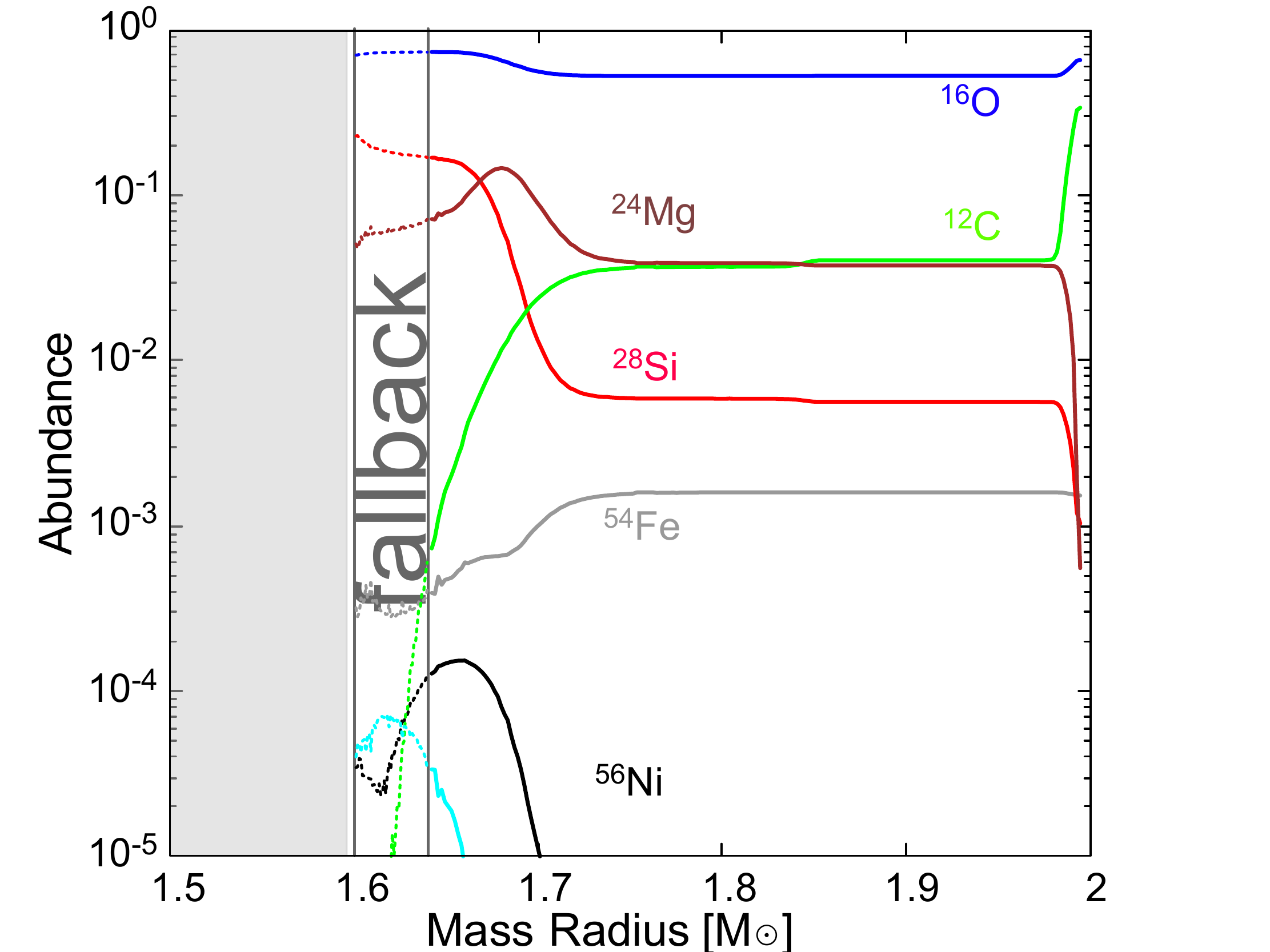} 
\caption{Same as Figure \ref{fig:CO145nuc} but for the CO20 model.} 
\label{fig:CO20nuc}
\end{figure}

Figs. \ref{fig:CO145nuc} and \ref{fig:CO20nuc} show results of our explosive nucleosynthesis calculation for the CO145 and CO20 models, respectively. 
In the lighter case, a few $\times 10^{-2}\,M_\odot$ of \nickel are synthesized during the explosion. A fraction of them falls back and $\sim 0.01\,M_\odot$ of \nickel are ejected. In the heavier case, although the amounts of \nickel synthesized are comparable, most of them fall back and only $\lesssim 10^{-4}\,M_\odot$ of \nickel are ejected. Table \ref{tbl:USSN} summarize the amount of $^{56}$Ni ejected in all cases.

Our results show that both the synthesized and ejected amounts of \nickel in the USSNe are significantly smaller than those estimated for canonical CCSNe. This trend is mainly due to the less compactness of the USSN progenitors and comparable proto-NS masses to canonical SN explosions. In more detail, this can be understood as follows. The synthesized amount of \nickel can be roughly calculated as the enclosed mass within the \nickel synthesizing radius $R_{T_9=5}$ (see Equation \ref{eq:app2}) minus the NS mass. A smaller explosion energy of USSNe gives a smaller $R_{T_9=5}$, and ultra-stripped progenitors have a smaller mass radius for a fixed radius than the canonical SN progenitor (see Figure \ref{fig:preSN}), thus USSNe have a smaller enclosed mass within the \nickel synthesizing radius. On the other hand, the calculated NS mass is roughly comparable between the USSNe and the canonical SNe. Resultantly, the synthesized amount of \nickel becomes smaller for the USSNe. The ejected amounts of \nickel also becomes smaller for the USSNe since the fallback masses and the proto-NS masses are comparable between the USSNe and the canonical SNe.

\begin{figure}
    \centering
    \includegraphics[width=0.475\textwidth]{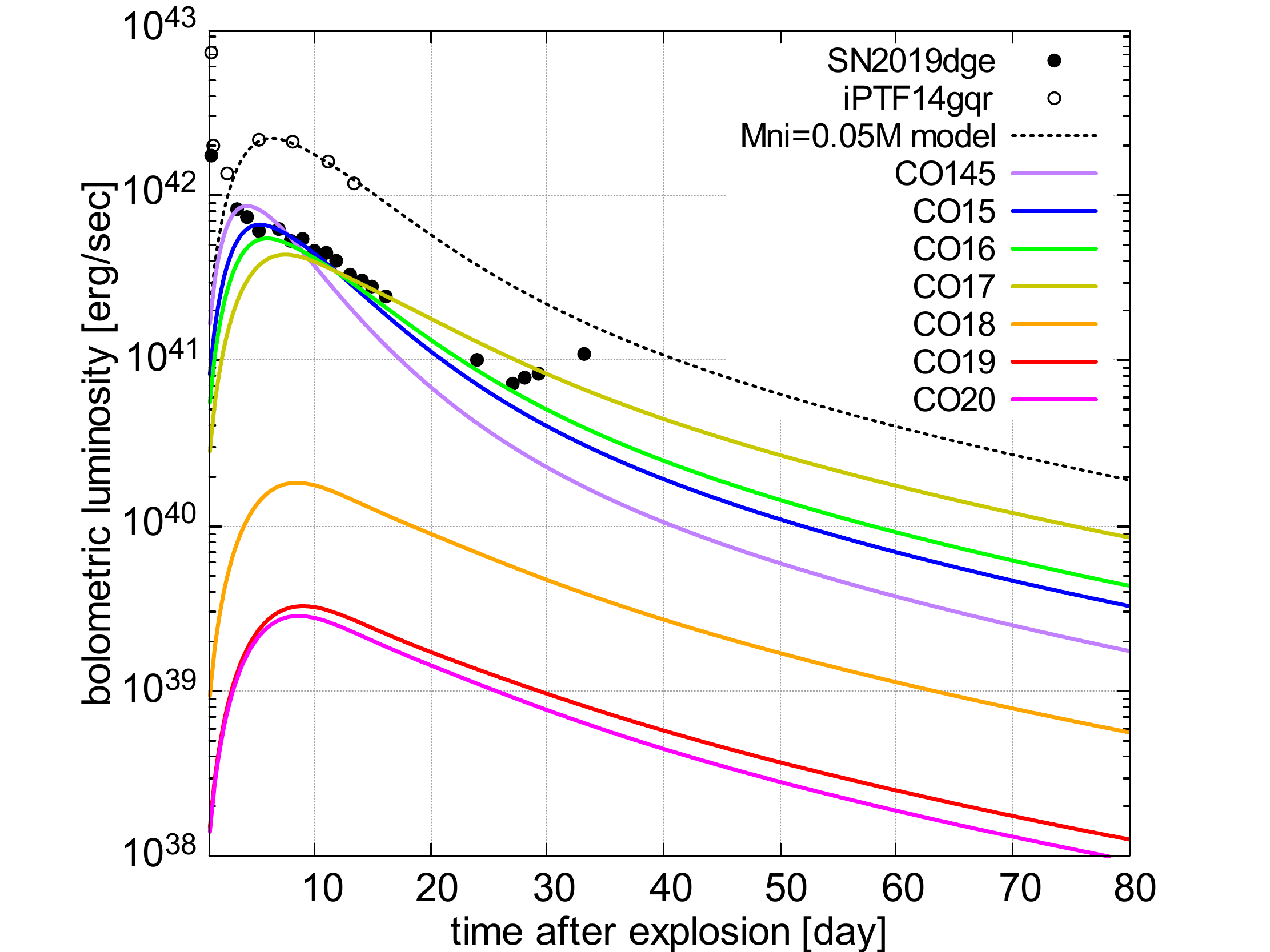} 
    \caption{The bolometric light curves of seven CO progenitor models.The empty and black circles are the estimated bolometric luminosity of iPTF 14gqr and SN 2019dge, respectively. 
    Also shown in the dotted line for comparison is the phenomenological best-fit model for iPTF 14gqr with $M_\mathrm{Ni} = 0.05M_\odot$, $M_\mathrm{ej} = 0.2M_\odot$, and $E_\mathrm{expl} = 0.2$ Bethe obtained in \citet{2018Sci...362..201D}} 
\label{fig:USSN-LC}
\end{figure}

Figure \ref{fig:USSN-LC} shows bolometric light curves of the USSNe. 
The ejecta mass $M_\mathrm{ej}$ and explosion energy $E_\mathrm{expl}$, among the input three parameters obtained by hydrodynamical simulations, determine the shape of the light curve, especially the peak time (see Eq \eqref{eq:rad_loss}).
When the ejecta mass $M_\mathrm{ej}$ is smaller, the peak of the light curve becomes faster. The peak also becomes faster when the explosion energy $E_\mathrm{expl}$ is larger.
Figure \ref{fig:USSN-LC} suggests that the differences in ejecta mass and explosion energy within the range of the present hydrodynamical results make little difference in the shape of the light curve.
And the other parameter, nickel mass $M_\mathrm{Ni}$, changes the peak luminosity.
Since an USSN from a more massive progenitor has a larger ejecta mass but a smaller \nickel mass, the light curve becomes dimmer and slower; the peak luminosity can be as large as $10^{42}$ erg s$^{-1}$ for the relatively small CO core mass cases where the ejected amount of \nickel is $M_\mathrm{Ni}\sim 0.01M_\odot$ while the peak luminosity is significantly smaller for the heaviest three cases (CO18, CO19, CO20) for which the ejected amount of \nickel is significantly small mainly due to the fallback. 

Figure \ref{fig:USSN-LC} also shows the observed bolometric light curves of iPTF 14gqr and SN 2019dge~\citep{2018Sci...362..201D,2020ApJ...900...46Y}. We find that SN 2019dge is broadly consistent with the relatively small CO core mass cases.\footnote{The observed light curve of SN 2019dge is reproduced by our CO16 model. The model parameters, especially the ejecta mass, are different from the best-fit parameters obtained in the previous work~\citep{2020ApJ...900...46Y}. This is mainly due to the different treatment of the $\gamma$-ray escape timescale in the light curve calculation (see Sec. \ref{sec:lc} in this paper and their Appendix B).}
However, none of our models can reproduce the peak luminosity of iPTF 14gqr. The dotted line indicates the best-fit model obtained by \cite{2018Sci...362..201D} with $M_\mathrm{Ni} = 0.05M_\odot$, $M_\mathrm{ej} = 0.2M_\odot$, and $E_\mathrm{expl} = 0.2$ Bethe. Such a large amount of \nickel cannot be ejected in our theoretical calculations. This inconsistency can be due to that (i) our model underestimates the ejected amount of \nickel in USSNe, or (ii) there is an additional energy source other than the \nickel decay at least for relatively bright USSNe. We consider both possibilities in the following section.

\section{Discussion}\label{sec:discus}

\subsection{$^{56}$Ni problem?}\label{sec:nickel}

Our long-term explosion simulation is based on the results of neutrino-hydrodynamics simulations in that the explosion energy $E_\mathrm{expl}$ and the initial proto-NS mass $M_\mathrm{PNS, i}$ are consistent. However, there are several model uncertainties which could potentially lead to underestimating the ejected amount of \nickel. We here investigate this point. 

\begin{figure}
    \centering
    \includegraphics[width=0.475\textwidth]{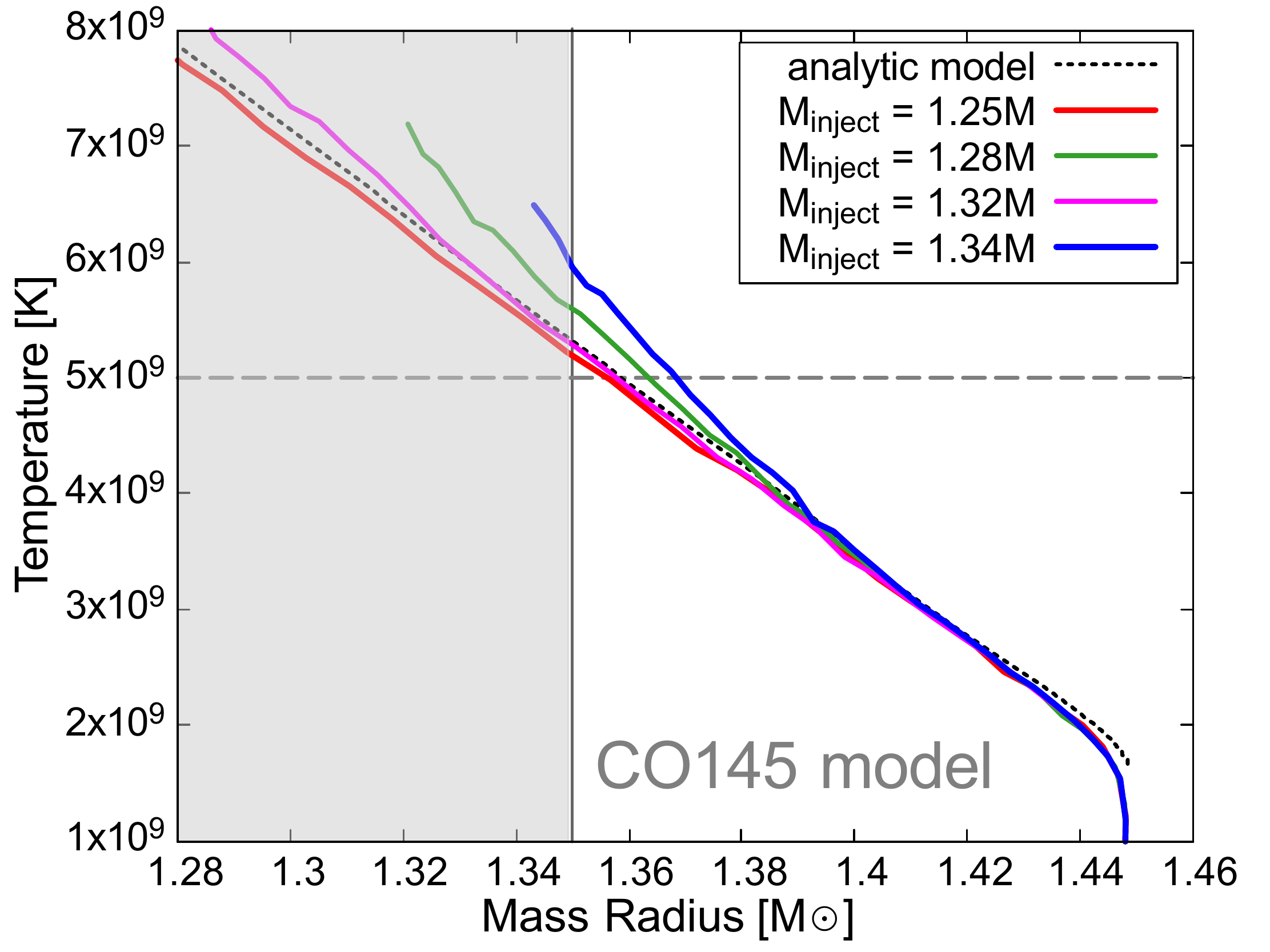} 
    \caption{The evolution of the temperature at the shock front of the CO145 model with various energy injection positions. The solid red line shows our fiducial case in Table \ref{tbl:USSN}, and the solid magenta, green and blue lines are the cases with $M_\mathrm{inject} \approx 1.28, 1.32$ and $1.34M_\odot$, respectively. The dashed black line is the analytic model given by Equation \eqref{eq:app1}. The gray region eventually becomes a neutron star.
    } \label{fig:CO145-check}
\end{figure}

First, we discuss the degeneracy in the explosion model : For a given progenitor model, a same ($E_\mathrm{expl}$, $M_\mathrm{PNS, i}$) can be obtained for a series of pair of injection energy and mass radius ($E_\mathrm{inject}$, $M_\mathrm{inject}$). Among them, we have employed the case with a sufficiently small $M_\mathrm{inject}$ that would better mimic the result of neutrino-hydrodynamics simulation. This choice is somewhat arbitrary and gives a conservative estimate on the synthesized \nickel mass. 
Figure \ref{fig:CO145-check} shows the evolution of the shock downstream temperature of the CO145 model with various $M_\mathrm{inject}$. For larger $M_\mathrm{inject}$ cases, the region experiencing temperatures above $5\times10^9$ K is more extended and thus more \nickel is produced.   
Nevertheless, Figure \ref{fig:CO145-check} indicates that the increase in the synthesized amount of \nickel compared with our fiducial case ($M_\mathrm{inject} = 1.25 M_\odot$) is at most $\sim0.02\,M_\odot$ for the CO145 model. Since the proto-NS masses from first-principles calculations is $1.35M_\odot$, the synthesizing radius $R_{T_9=5}$ (see Equation \ref{eq:app2}) would have to extend to around $1.4M_\odot$ to synthesize $\sim 0.05M_\odot$ of $^{56}$Ni, which is very difficult to achieve. We also confirm this is also the case for heavier CO core models.

\begin{figure}
    \centering
    \includegraphics[width=0.475\textwidth]{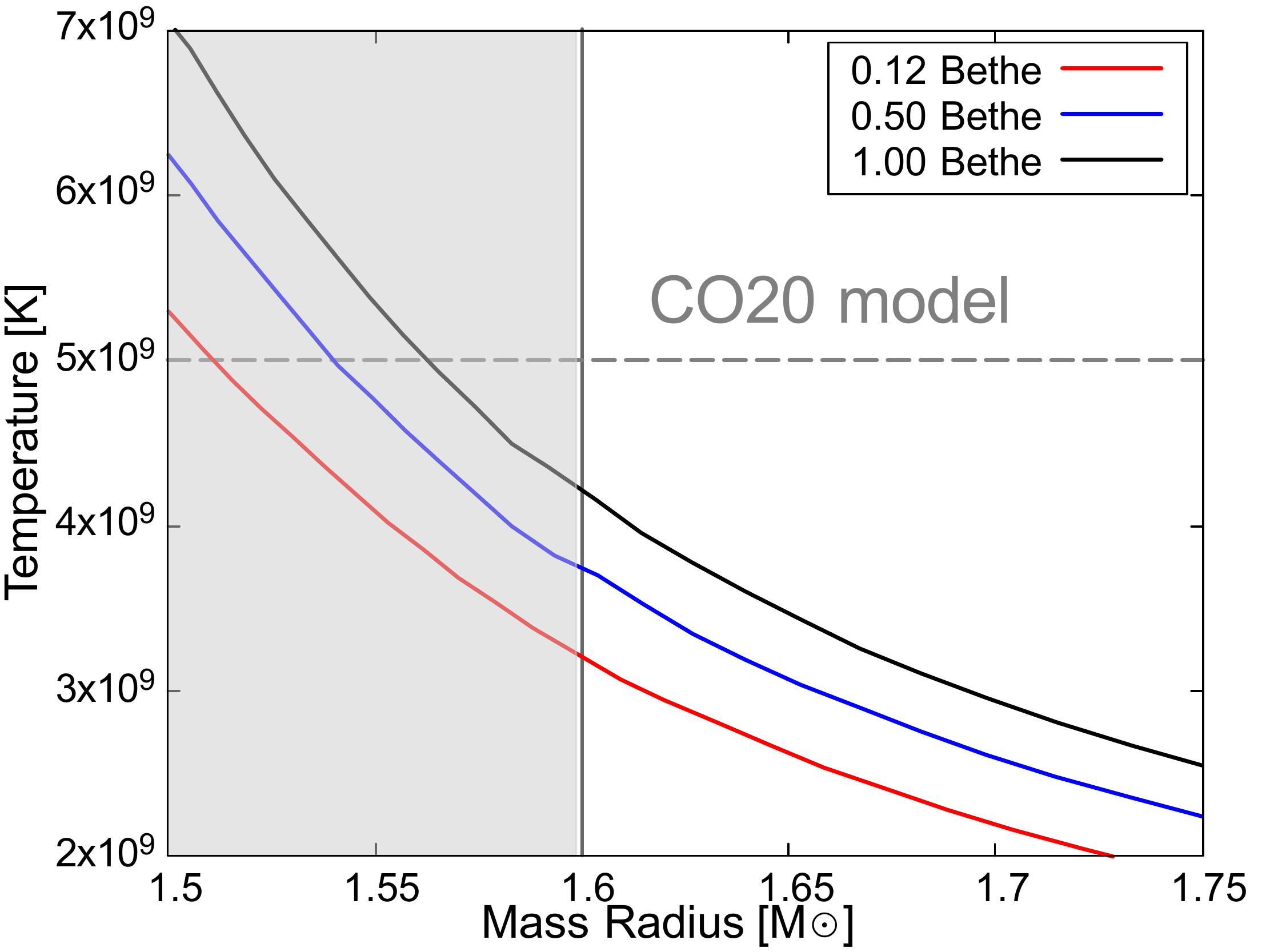} 
    \caption{The evolution of the temperature at the shock front of the CO20 model with various explosion energies. The solid red line shows our fiducial case in Table \ref{tbl:USSN}, and the blue and black lines are the cases with $E_\mathrm{expl.}\approx 0.5$ and $1.0$ Bethe respectively. The gray region eventually becomes a neutron star.
    } 
    \label{fig:CO20-check}
\end{figure}

Next, we discuss the uncertainties of the explosion energy in massive progenitor models.
$E_\mathrm{expl}= 0.12$ Bethe adopted for heavier than CO16 models is the only assumption, since the final explosion energy is not obtained by the neutrino-radiation hydrodynamics simulations due to their weak explosions and the limited simulation period.
We here investigate how the explosion energy would affect the hydrodynamic behavior and nucleosynthesis.
Figure \ref{fig:CO20-check} shows the evolution of the peak temperature behind the shock wave for different explosion energies.
We confirm that while the nickel synthesis region is extended by about $0.06 M_\odot$ due to larger energy, all of them are still entirely contained within the fallback region and are not expected to escape from the central proto-NS.

Other possible indeterminacy include neutrino-driven wind, which is expected to produce additional \nickel after explosive synthesis. 
\cite{2021ApJ...908....6S} claimed that the \nickel amount by neutrino-driven wind is strongly related to the mass accretion rate onto the proto-NS just after the onset of the explosion. For the case of USSN that has a small envelope mass and negligible mass accretion during the important phase, the neutrino-driven wind is not expected to fill the gap described above.

In summary, even taking into account the model uncertainties, our calculations indicate that it is difficult to synthesize $\sim 0.05M_\odot$ of $^{56}$Ni in an USSN, which may conflict with the observed light curves of SN 2005ek and iPTF 14gqr.
First-principle simulations of the entire USSN explosion and more involved calculations of the ultra-stripped progenitors including various effects of binary evolution will be needed to further investigate this issue.

\subsection{Alternative energy sources}\label{sec:obs}
Here we consider alternative energy sources of USSNe other than \nickel decay.
One possibility is the fallback accretion onto the newborn NS, e.g., a fraction of the gravitational energy of the accreted matter can be released via outflow from an accretion disk formed by the fallback materials~\citep[e.g.,][]{2012MNRAS.423.3083M,2013ApJ...772...30D}. Another possibility is the pulsar wind, i.e. the rotational energy of the newborn NS is extracted by the dipole radiation and forms a nascent pulsar wind nebula~\citep[e.g.,][]{2017ApJ...850...18H}. We calculate the USSN light curves powered by these additional energy sources and compare them in particular with iPTF 14gqr.

\subsubsection{Fallback accretion}
To include the additional energy injection to the SN ejecta by the fallback accretion, we modify the heating term in Equation \eqref{eq:ap14} as follows~\citep[e.g.,][]{2013ApJ...772...30D,2018ApJ...867..113M};
\begin{align}
    &\dot{Q}(t) =  f_\mathrm{dep}
    \left(  M_\mathrm{Ni} \,\ q_\mathrm{Ni}(t) \right)
    +\dot{Q}_\mathrm{acc}(t),\\
    &\dot{Q}_\mathrm{acc}(t)=\eta\dot{M}_\mathrm{fb}(t) c^2 \label{eq:lc2}~,
\end{align}
where $\dot{M}_\mathrm{fb}(t)$ is the fallback rate obtained by our simulation (see Equation \eqref{eq:basic6} and Figure \ref{fig:fbrate}), and $\eta$ is the energy conversion efficiency, which is largely uncertain. Here we consider a range of $\eta \leq 10^{-3}$ following \citet{2013ApJ...772...30D}.

\begin{figure*}
\gridline{\fig{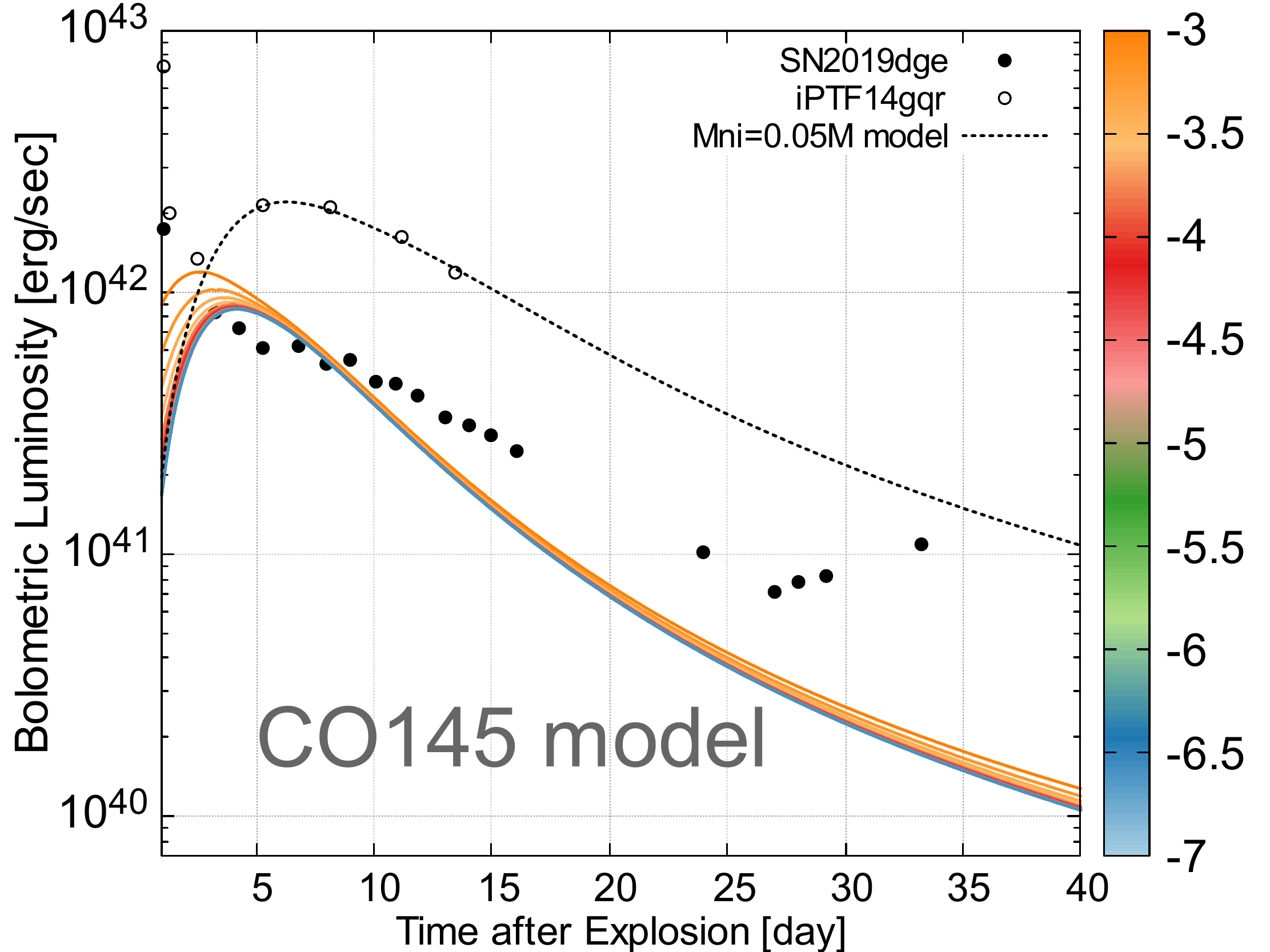}{0.333
\textwidth}{(a)CO145 model}
          \fig{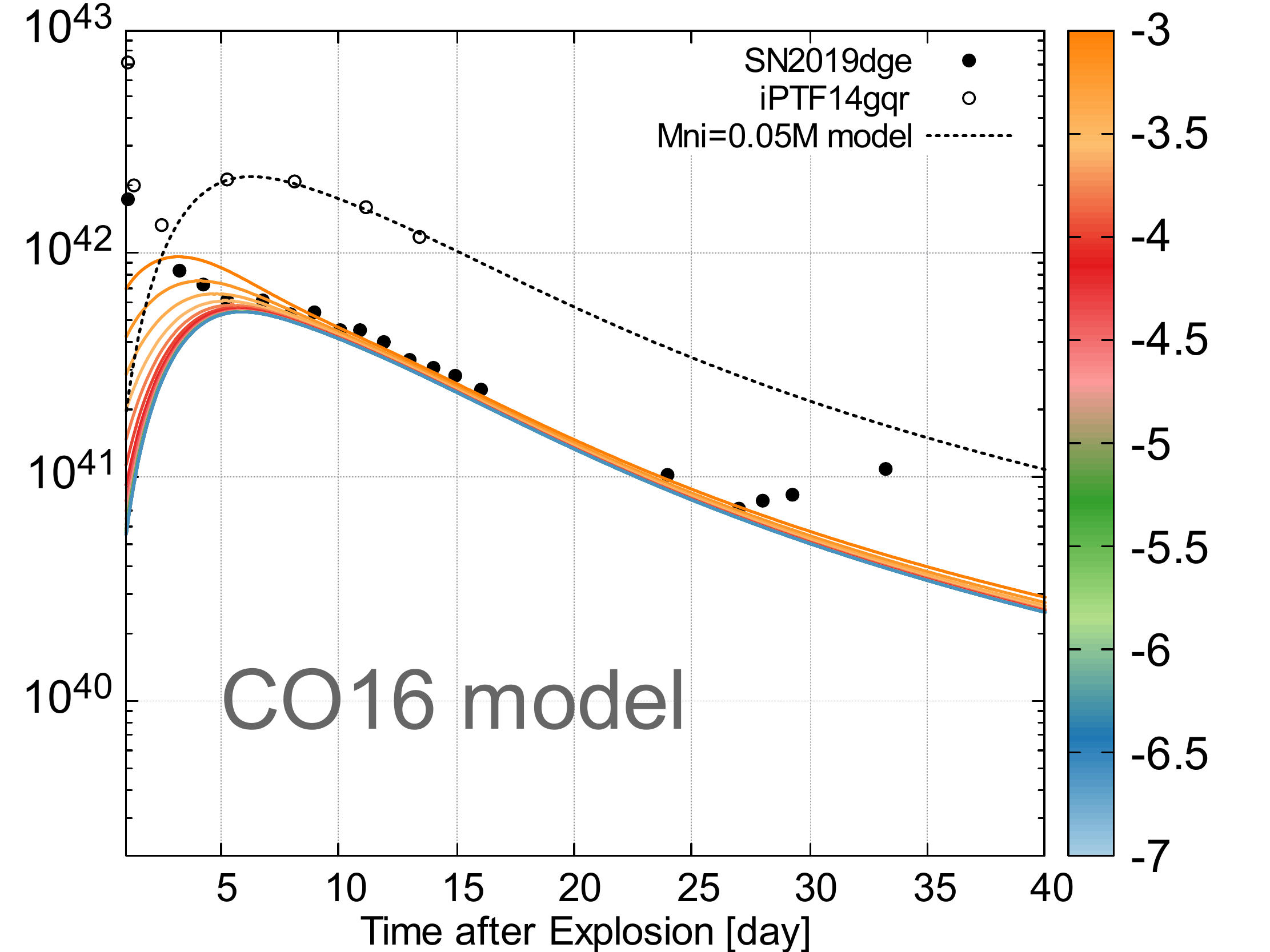}{0.333\textwidth}{(b)CO16 model}
          \fig{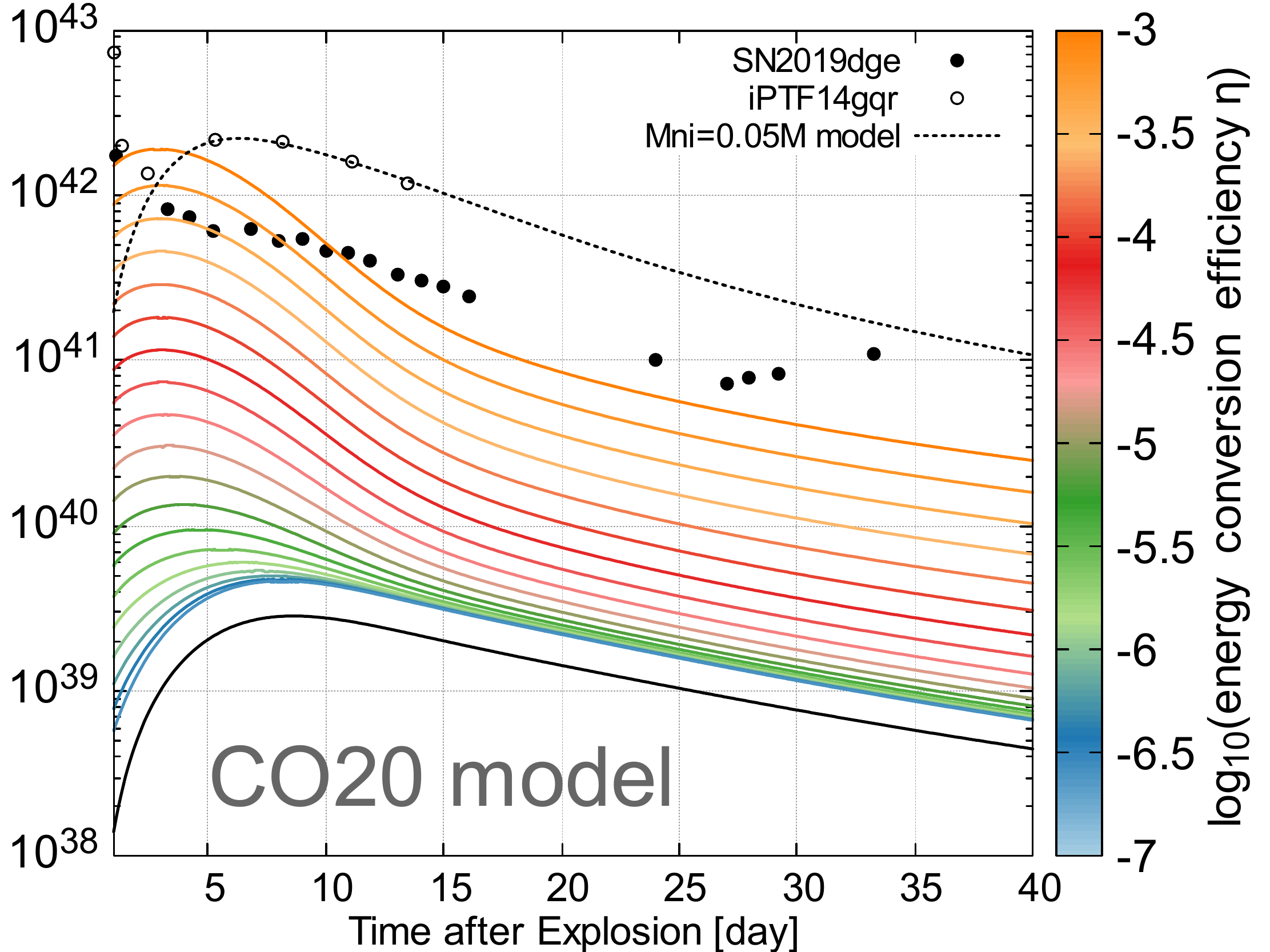}{0.333\textwidth}{(c)CO20 model}
          }
    \caption{The bolometric light curves of CO145, CO16 and CO20 models with fallback-heating.
    The color bar shows the logarithmic value of the energy conversion efficiency $\eta$.
    The black lines are models with pure \nickel-decay heating.
    Also shown in the dotted line for comparison is the same phenomenological best-fit model as in Figure \ref{fig:USSN-LC}} 
\label{fig:USSN-LC-acc}
\end{figure*}

Figure \ref{fig:USSN-LC-acc} shows the bolometric light curves of the CO145, CO 16, and CO20 models including the energy injection by the fallback accretion. We find that the fallback accretion can make the USSN light curve brighter and faster. In particular, the fallback accretion can significantly increase the peak luminosity of the CO20 model where the ejected amount of \nickel is suppressed due to the intense fallback itself. Even in this case, however, the light curve shape is incompatible with iPTF 14gqr. Since the fallback accretion rate decreases faster than the \nickel decay rate, it mainly contributes to the very early phase ($t \lesssim 5\,\mathrm{day}$). This effect is specific to the USSN, which shows a fast-rising light curve due to its extremely low ejecta mass. Thus, we also confirmed that there is no significant effect when adapting similar energy injection to canonical SNe in the range of typical fallback accretion.
We conclude that the fallback accretion could be an important energy source for USSNe especially in the very early phase, but could not be the main energy source for relatively bright USSNe like iPTF 14gqr.

\subsubsection{Rotation-powered relativistic wind}
In order to incorporate the energy injection by the rotation-powered relativistic wind to the light curve calculation, we modify the heating term in Eqs (\ref{eq:ap1}) and (\ref{eq:ap2}) as 
\begin{align}
    &\dot{Q}(t) = f_\mathrm{dep}
    \left(  M_\mathrm{Ni} \,\ q_\mathrm{Ni}(t) \right)
    +\dot{Q}_\mathrm{p}(t)\\
    &\dot{Q}_\mathrm{p}(t)=\cfrac{E_p}{t_p}\cfrac{1}{(1+t/t_p)^2} \label{eq:lc3}~,
\end{align}
where $E_p = 2\times10^{50}\,P_{10}^{-2}$ erg is the rotational energy, $t_p = 0.44\,B_{14}^{-2} P_{10}^{2}$ yr is the spin-down timescale, $B_{14}=(B/10^{14}\,\mathrm{G})$ is the dipole magnetic field strength, and $P_{10}=(P_i/10\,\mathrm{ms})$ is the spin period of the newborn NS ~\citep[]{2010ApJ...717..245K}. The NS radius is set to be $R_\mathrm{ns} = 12$ km. In Eq. (\ref{eq:lc3}), we assume the energy conversion efficiency from the rotation-powered wind to the heat as $\mathcal{O}(1)$, which is appropriate for the bolometric light curve at around the peak~\citep{2017ApJ...850...18H}, and also neglect the spin down of the NS via gravitational-wave radiation~\citep[e.g.,][]{2016ApJ...818...94K}. Such wind-driven USSN light curves can have diverse shapes depending on $B$ and $P_i$~\citep{2017ApJ...850...18H}. Here we focus on searching the parameter set that can reproduce the light curve of iPTF 14gqr.

\begin{figure}
    \centering
    \includegraphics[width=0.475\textwidth]{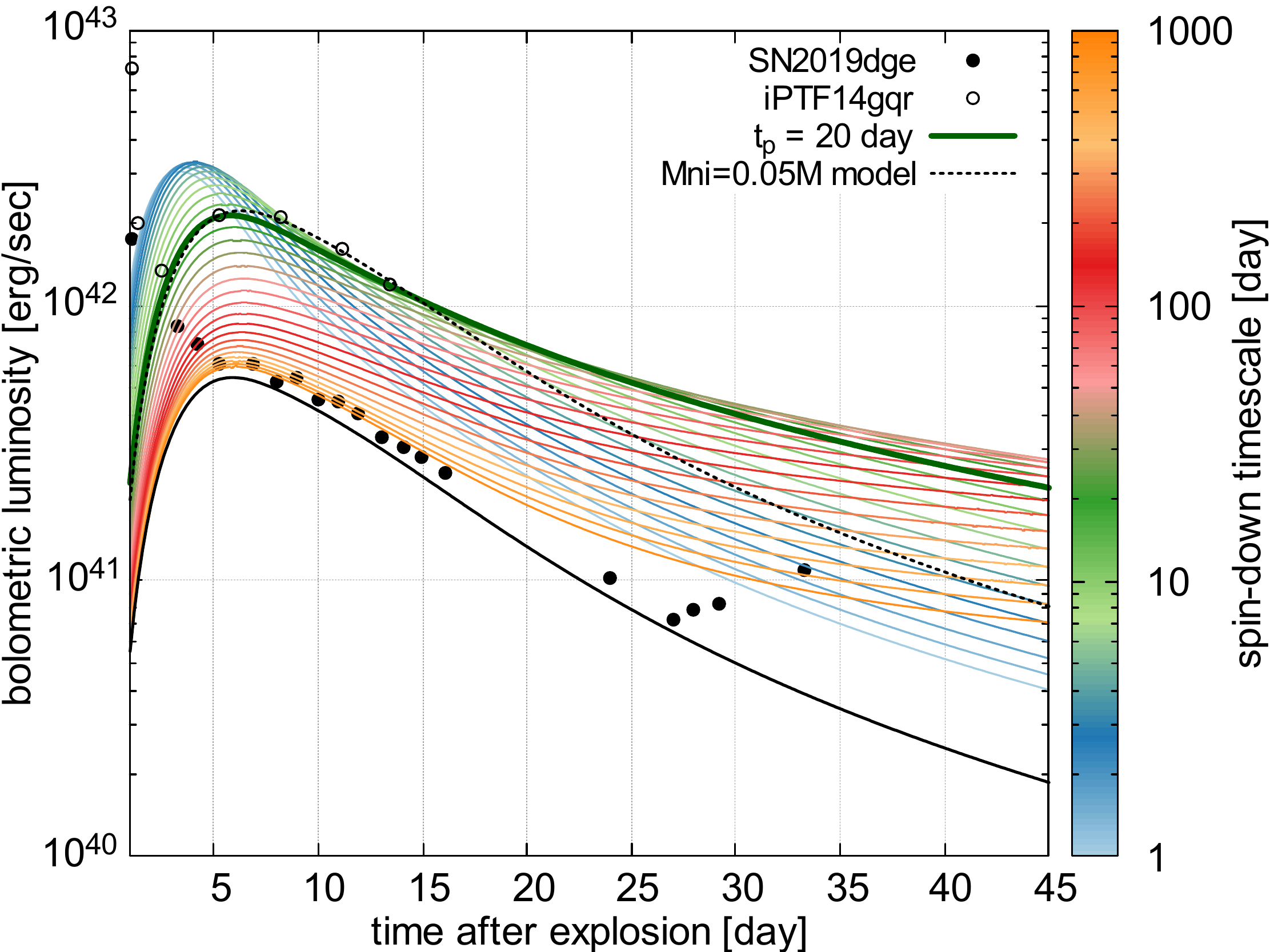}
    \caption{The bolometric light curves of CO16 models with pulsar wind energy. 
    The color bar shows the spin-down timescale $t_p$, and $E_p=4.0\times10^{48}$ erg is fixed.
    The black lines are models with pure \nickel-decay heating.
    The best fit bolometric light curve models of iPTF 14gqr is marked by thick line.
    Also shown in the dotted line for comparison is the same phenomenological best-fit model as in Figure \ref{fig:USSN-LC}} 
\label{fig:USSN-LC-pulsar}
\end{figure}

Figure \ref{fig:USSN-LC-pulsar} shows the bolometric light curves of the CO16 model with the additional energy injection by the rotation-powered wind. 
We investigate a parameter region of $1.0\times 10^{48}\,\mathrm{erg} < E_p < 4.0 \times 10^{50} \,\ \mathrm{erg}$ and $1\,\mathrm{day} < t_p < 1000\,\mathrm{day}$, and show the cases with an fixed rotation energy ($E_p = 4.0\times 10^{48}\,\mathrm{erg}$) for representative purpose. We find that the bolometric light curve of iPTF 14gqr can be well fitted with $E_p = 4.0\times 10^{48}\,\mathrm{erg}$ and $t_p = 19.8\,\mathrm{day}$, namely $B \approx 2.0\times10^{15}\,\mathrm{G}$ and $P_i \approx 70\,\mathrm{ms}$. A similar set of fitting parameter ($B$,$P_i$) is obtained when other CO core mass models are used; the light-curve shape is still characterized by the diffusion timescale of the USSN ejecta, which is roughly comparable in the original models (see section \ref{sec:result}), while the peak luminosity is mainly determined by $\sim E_p/t_p$ or ($B$,$P_i$).

We note that the above NS parameters for iPTF 14gqr are compatible with the Galactic magnetars~\citep[e.g.,][]{enoto+19}, but different from those inferred for superluminous supernovae ~\citep[$B \sim 10^{13\mbox{-}14}\,\mathrm{G}$ and $P_i \sim 1\mbox{-}10\,\mathrm{ms}$; e.g.,][]{2016ApJ...818...94K}. 
Our result might imply that such a magnetar formation is relatively common in core-collapse explosions in a close binary system. 

\section{Conclusions}\label{sec:concl}

We have conducted the explosion simulations of ultra-stripped progenitors with various CO-core masses based on the results of neutrino-radiation hydrodynamics simulations, and consistently calculate the nucleosynthesis and the SN light curves.
We found the amount of $^{56}$Ni synthesized and ejected in the USSNe are smaller than those in canonical CCSNe mainly due to the less compactness of the USSN progenitors and comparable proto-NS masses to canonical CCSNe.
Nonetheless, the relatively small CO-core mass models can eject sufficient amounts of $^{56}$Ni to reproduce the light curves of some observed USSNe \citep[e.g., SN2019dge;][]{2020ApJ...900...46Y}.
On the other hand, it is difficult for any progenitor model to synthesize $\sim 0.05M_\odot$ of $^{56}$Ni as inferred for relatively bright USSNe \citep[e.g., SN 2005ek and iPTF 14gqr;][]{2013ApJ...778L..23T,2018Sci...362..201D}.
We have investigated the fallback accretion onto and relativistic wind from the newborn NS as alternative energy sources for such USSNe. 
We found that iPTF 14gqr can be explained by considering an energy injection from a newborn NS with a magnetic field of $B_p \sim 10^{15}\,\mathrm{G}$ and an initial rotation period of $P_i \sim 0.1\,\mathrm{s}$.

\software{{\tt MESA} \citep{2011ApJS..192....3P,2013ApJS..208....4P,2015ApJS..220...15P,2018ApJS..234...34P,2019ApJS..243...10P}, 
{\tt torch} \citep{1999ApJS..124..241T}, 
{\tt hydro1d} (https://zingale.github.io/hydro1d/)}

\section*{Acknowledgments}
We thank T. Yoshida for providing progenitor models.
The work has been supported by Japan Society for the
Promotion of Science (JSPS) KAKENHI grants 
21J00825, 21K13964 (R.S.), 20K04010 (K.K), 18H05437, 20H00174, 20H01904, and 20H04747 (Y.S.).

\restartappendixnumbering
\appendix
\section{ Maximum amount of $^{56}$Ni in the thermal bomb explosion model}\label{appendix}
Here, as a supplement to the discussion of the degeneracy of the explosion model in Section \ref{sec:nickel}, we show the maximum amount of \nickel that can be synthesized by the explosion model with a thermal bomb.
We find that the highest amount of synthesized \nickel is given by the model with the outermost energy injection position.
It allows us to discuss the most robust maximum limit on the amount of \nickel that can be ejected by the USSNe, within the range of reproducing the same ($E_\mathrm{expl}$, $M_\mathrm{PNS, i}$) as neutrino-radiation hydrodynamics simulations.
Note, however, that this setup does not allow us to estimate the natural fallback mass accretion rate, so we do not treat this choice as the primary model in this study.

\begin{figure}
\centering
  \includegraphics[width=0.475\textwidth]{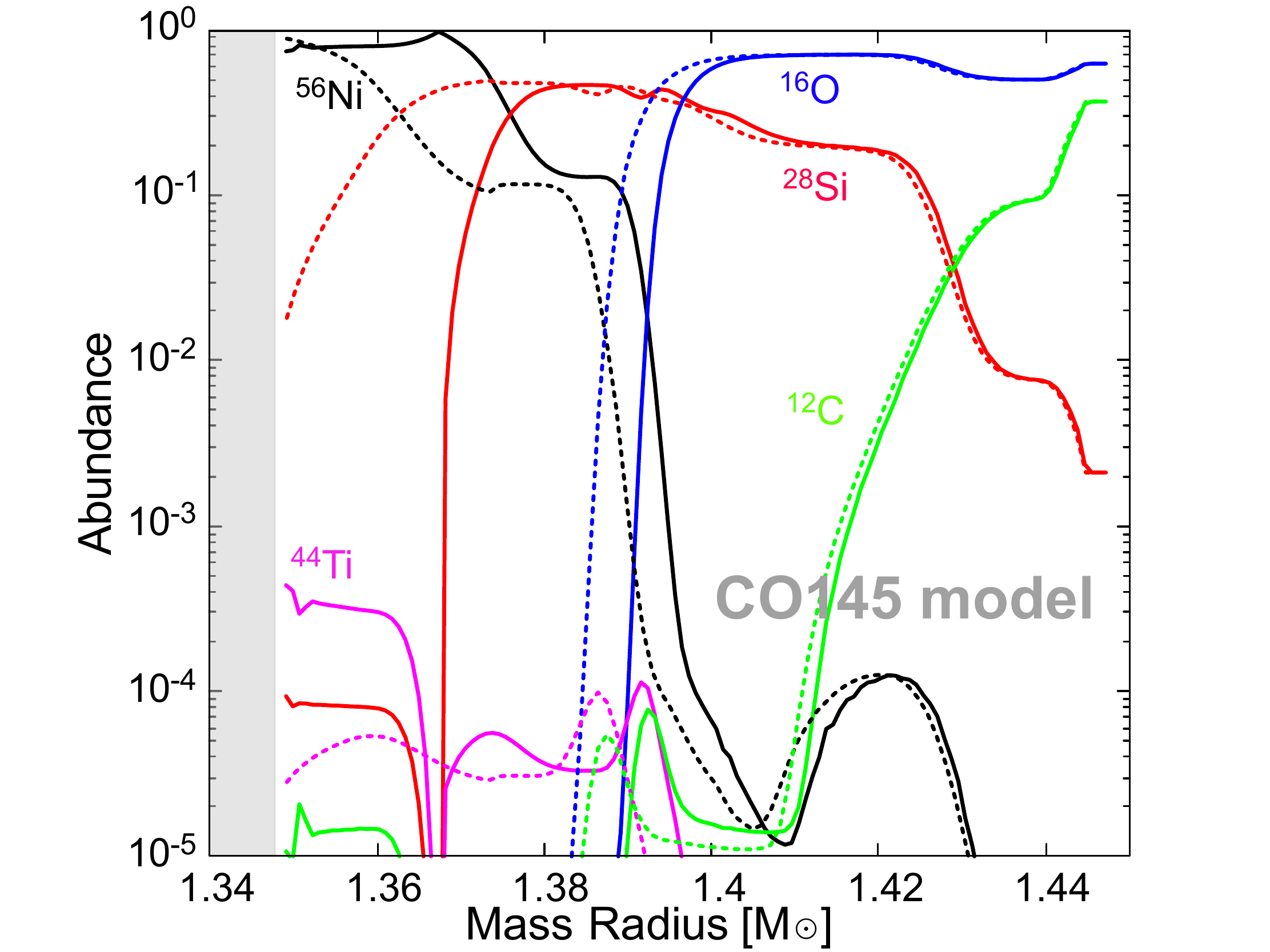} 
  \includegraphics[width=0.475\textwidth]{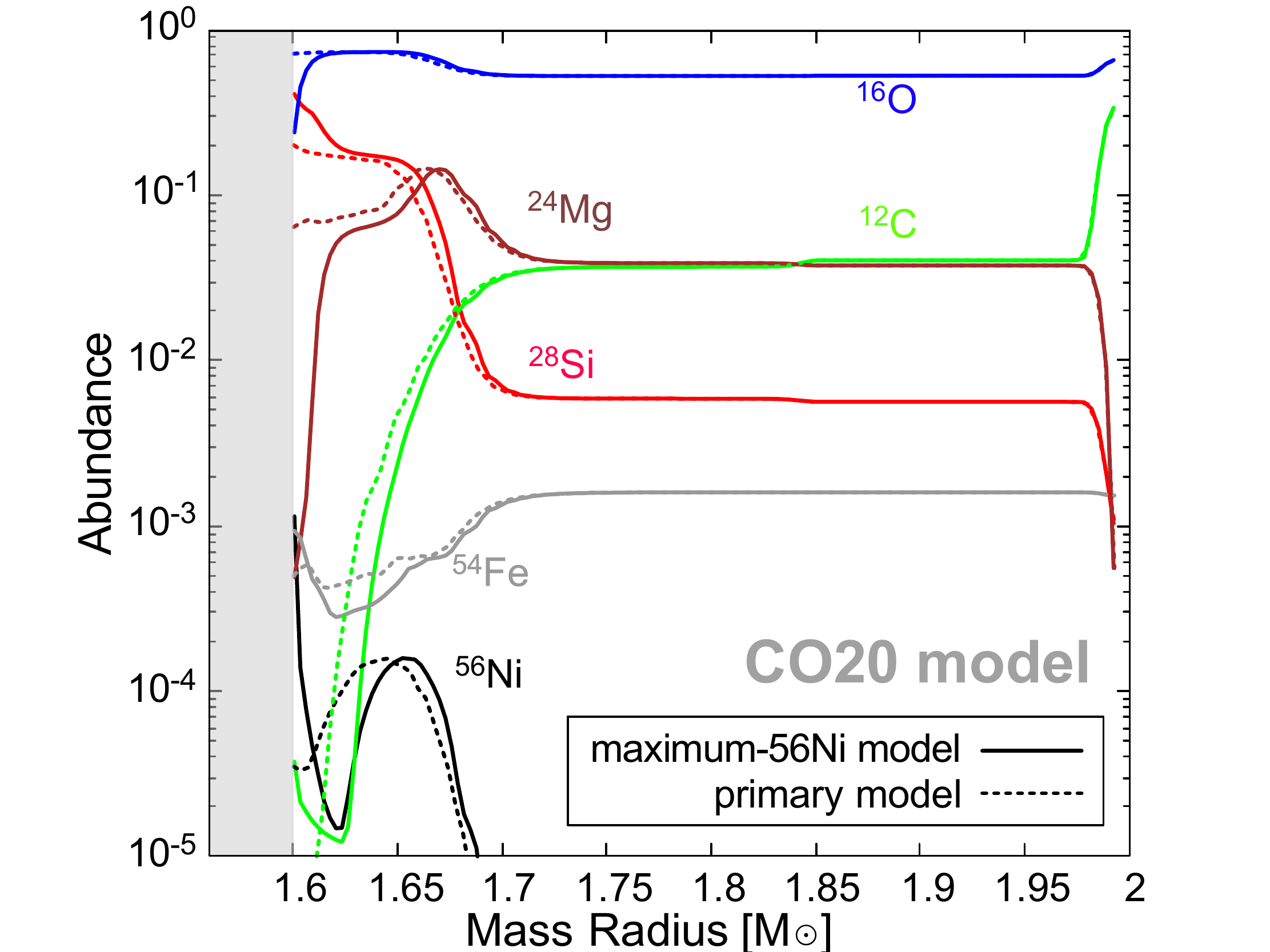} 
\caption{The results of Explosive nucleosynthesis in the CO145 and CO20 models of maximum \nickel synthesis setup, compared to the primary setup.}
\label{fig:appnuc}
\end{figure}

Figure \ref{fig:appnuc} shows the results of our explosive nucleosynthesis calculation with the outermost energy injection position for the CO145 and CO20 models, respectively. 
CO145 is the model with the highest amount of \nickel synthesis, while CO20 is the model with the lowest amount. The results of other models with similar setups are summarized in Table \ref{tbl:app}.
\cite{2018ApJ...867..113M} adopted the same thermal bomb injection setup as the current one, so our results are consistent with their results, which are $M_\mathrm{Ni}\sim0.026$ $M_\odot$ at similar explosion energies  ($E_\mathrm{expl}=0.10$ Bethe) and CO star masses ($M_\mathrm{CO}=1.50~M_\odot$). 
In order to reproduce the explosion energy of $\sim0.2$ Bethe and synthesized \nickel mass of $\sim0.05$ $M_\odot$ estimated for iPTF14gqr, the explosive nucleosynthesis would have to proceed up to $\sim 1.40$ $M_\odot$, for instance, in the CO145 model. Figure \ref{fig:appnuc} shows that it is very difficult to achieve.
We emphasize again that this result is based on a model that leads to the maximum amount of \nickel synthesis, and then we can give a more robust conclusion that it is difficult to synthesize $\sim 0.05M_\odot$ of $^{56}$Ni in an USSN.

\begin{table*}
\centering
    \caption{Summary of Explosive nucleosynthesis results with maximum \nickel synthesis setup.}\label{tbl:app}
    \begin{tabular}{ r  c  c c || c | c } \hline \hline
    &\multicolumn{3}{c||}{ Our work} &
    primary result  & maximum-\nickel result \\\hline 
      Model & $M_\mathrm{CO}$ 
      & $E_\mathrm{expl}$ & $M_\mathrm{PNS,i}$ 
      & $M_\mathrm{Ni}$ 
      & $M_\mathrm{Ni}$  \\
      &  ($M_\odot$)& (Bethe) & ($M_\odot$) & ($M_\odot$) & ($M_\odot$) \\ \hline 
      CO145& 1.45 & 0.17 & 1.35 & $1.63\times10^{-2}$ & 
      $2.88\times10^{-2}$ \\
      CO15 & 1.5  & 0.15 & 1.36 & $1.38\times10^{-2}$ &  $2.84\times10^{-2}$\\
      CO16 & 1.6  & 0.12 & 1.42 & $1.20\times10^{-2}$ &  $2.56\times10^{-2}$\\
      CO17 & 1.7  & 0.12 & 1.45 & $1.09\times10^{-2}$ &  $1.98\times10^{-2}$\\
      CO18 & 1.8  & 0.12 & 1.49 & $4.80\times10^{-4}$ &  $5.62\times10^{-3}$\\
      CO19 & 1.9  & 0.12 & 1.54 & $9.19\times10^{-5}$ &  $2.06\times10^{-4}$\\
      CO20 & 2.0  & 0.12 & 1.60 & $7.78\times10^{-5}$ &  $4.31\times10^{-4}$ \\
      \hline \hline
    \end{tabular}
\end{table*}

\bibliography{ref}{}
\bibliographystyle{aasjournal}

\end{document}